\documentclass[11pt]{article}

\usepackage{amsthm,amsmath,natbib}
\usepackage[blocks]{authblk}
\usepackage{parskip}
\usepackage{amsfonts} 
\usepackage{enumerate}
\usepackage{graphicx}
\usepackage{rotating}
\usepackage{natbib}

\newtheorem{prop}{Proposition}[section]

\newtheorem{theorem}{Theorem}[section]
\newtheorem{remark}{Remark}[section]

\newcommand{\deriv}{\frac{\partial}{\partial \Sigma}}
\newcommand{\F}{\mathcal{F}}

\newcommand{\E}{\mathbb{E}}

\newcommand{\D}{\mathcal{D}}
\newcommand{\HH}{\mathcal{H}}
\newcommand{\argmin}{\mathop{\arg \min}\limits}
\newcommand{\argmax}{\mathop{\arg \max}\limits}
\newcommand{\abs}[1]{\left| #1 \right|}

\newcommand{\calN}{\mathcal{N}}
\newcommand{\calT}{\mathcal{T}}
\newcommand{\X}{\mathbf{X}}
\newcommand{\PP}{\mathrm{P}}
\newcommand{\ZZ}{\mathrm{Z}}

\begin{document}
\title{Pattern Alternating Maximization Algorithm for Missing Data in ``Large P, Small N'' Problems}

\author[1]{Nicolas St\"adler}
\author[2]{Daniel J. Stekhoven}
\author[2]{Peter B\"uhlmann}
\affil[1]{\small The Netherlands Cancer Institute, 1066 CX Amsterdam, The
  Netherlands.}
\affil[2]{\small Seminar f\"ur Statistik, ETH Z\"urich, CH-8092 Z\"urich,
  Switzerland.}

\maketitle

\begin{abstract}
We propose a new and computationally efficient algorithm for maximizing the
observed log-likelihood for a multivariate normal data matrix with missing
values. We show that our procedure based on iteratively regressing the
missing on the observed variables, generalizes the standard EM
algorithm by alternating between different complete data spaces and
performing the E-Step incrementally. In this non-standard setup we prove
numerical convergence to a stationary point
of the observed log-likelihood.

For high-dimensional data, where the number of variables may greatly exceed
sample size, we add a Lasso penalty in the regression part of our algorithm
and perform coordinate descent approximations. This leads to a
computationally very attractive technique with sparse regression
coefficients for missing data imputation. Simulations and results on four
microarray datasets show that the new method often outperforms other imputation techniques as k-nearest neighbors imputation, nuclear norm
minimization or a penalized likelihood approach with an $\ell_1$-penalty on
the inverse covariance matrix.
\vspace{0.5cm} 
\end{abstract}

\newpage
\section{Introduction and Motivation} 
Missing data imputation for large datasets is an essential pre-processing
step in complex data applications. A well-known example are microarray datasets which
contain the expression values of thousands of genes from a series
of experiments. Missing values are inherent
to such datasets. They occur for diverse reasons, e.g. insufficient
resolution, image corruption, dust or scratches on the slides. Apart from microarrays, much
attention has been recently given to the so-called matrix completion
problem. Most prominent in this
context is the ``Netflix''
movie rating dataset with rows corresponding to customers and columns
representing their movie rating and the customers have seen/rated only a small
fraction of all possible movies. The goal is to estimate the ratings
for unseen movies. Filling in missing values in gene
  expression data is a inherently different problem from dealing with
  missings in the context of matrix completion. For example the
  Netflix dataset involves a  huge number of customers (480'000) and
  movies (17'000) with about 98\% of the ratings missing. In big
  contrast, microarrays have the typical ``large p, small n'' form where
  the number of different genes $p$ is in the ten thousands and the number of
  experimental conditions $n$ is in the hundreds. Usually only a small fraction of the expression values are missing. In this
paper, we propose a new and computationally efficient EM-type algorithm for missing value imputation in the
high-dimensional multivariate normal model where the number of
variables $p$ can greatly exceed the number of independent samples
$n$. Our motivating examples are microarray datasets with
  $p$ different genes (variables) and $n$ different experimental conditions
(samples).
The Gaussian assumption in our model is used for
computation of the likelihood but empirical findings suggest that the
method is useful for many continuous data matrices.

There is a growing literature of missing value imputation methods. For
microarray data, examples are k-nearest neighbors imputation and singular
value decomposition imputation \citep{knnimpute2001}, imputation based on
Bayesian principal component analysis \citep{bpcaimpute} or the local least
squares imputation from \cite{localleastsquares2006}. For a review and a
discussion on microarray data imputation see \cite{aittokallio2009}. In the
context of the so-called matrix completion problem, where the goal is to
recover a low-rank matrix from an incomplete set of entries, it has been
shown in a series of fascinating papers that one can recover the missing
data entries by solving a convex optimization problem, namely, nuclear-norm
minimization subject to data constraints
\citep{candesandrecht2009,candesandtao2009,keshavan2009}. Efficient convex
algorithms for the matrix completion problem were proposed by
\cite{svt2008} and \cite{softimpute2008}. If the missing data problem does
not arise from a near low rank matrix scenario, there is substantial room
to improve upon the convex matrix completion algorithms. We will
empirically demonstrate this point for some microarray high-throughput
biological data.





Here, we address the missing data problem through a likelihood approach
\citep{LittleRubin,Schafer}. We model the correlation between different
variables in the data by using the Multivariate Normal Model (MVN)
$\calN(\mu,\Sigma)$ with a p-dimensional covariance matrix
$\Sigma$. Recently, in the high-dimensional setup with $p\gg n$,
\cite{missglasso2010} proposed to maximize the penalized observed
log-likelihood with an $\ell_1$-penalty on the inverse covariance
matrix. They called their method \emph{MissGLasso}, as an extension of
the GLasso \citep{friedman2007sic} for missing data. A similar approach, in
the context of so-called transposable models, is given by
\cite{transposable2010}.

The \emph{MissGLasso} uses an EM algorithm for optimization of the
penalized observed log-likelihood. Roughly, the algorithm can be summarized
as follows. In the E-Step, for each sample, the regression coefficients of
the missing against the observed variables are computed from the current
estimated covariance matrix $\hat{\Sigma}$. In the following M-Step, the
missing values are imputed by linear regressions and $\hat{\Sigma}$ is
re-estimated by performing a GLasso on completed data. There are two main
drawbacks of this algorithm in a high-dimensional context. First, the
E-Step is rather complex as it involves (for each sample) inversion and
multiplication of large matrices in order to compute the regression
coefficients. Secondly, a sparse inverse covariance does not imply sparse
regression coefficients while we believe that in high-dimensions, sparse
regression coefficients would enhance imputations.

Our new algorithm, \emph{MissPALasso} (Missingness Pattern Alternating
Lasso algorithm) in this paper, generalizes the E-Step in order to resolve
the disadvantages of \emph{MissGLasso}. In particular, inversion of a
matrix (in order to compute the regression coefficients) will be replaced
by a simple soft-thresholding operator. In addition, the regression
coefficients will be sparse, which leads to a new sparsity concept for
missing data estimation.

In order to motivate \emph{MissPALasso}, we develop first the
Missingness Pattern Alternating maximization algorithm (\emph{MissPA}) for
optimizing the (unpenalized) observed log-likelihood. \emph{MissPA}
generalizes the E- \emph{and} M-Step of the standard EM originally
proposed by \cite{Dempster} by alternating between different complete data
spaces \emph{and} performing the E-Step incrementally. Such a
generalization does not fit into any of the existing methodologies which
extend the standard
EM. 
However, by exploiting the special structure of our procedure and applying
standard properties of the Kullback-Leibler divergence, we prove
convergence to a stationary point of the observed log-likelihood.

The further organization of the paper is as follows: Section 2 introduces
the setup and the useful notion of missingness patterns. In Section 3 we
develop our algorithms: \emph{MissPA} is presented in Section~\ref{sec:patternwise} and \emph{MissPALasso} then follows in Section~\ref{sec:patternwiselasso} as an
adapted version for high-dimensional data with $p\gg n$. Section 4 compares
performance of \emph{MissPALasso} with other imputation methods on
simulated and real data and reports on computational efficiency. Finally,
in Section 5, we present some mathematical theory which describes the
numerical properties of the Missingness Pattern Alternating maximization algorithm.
 
\section{Setup}\label{sec:setup}
We assume 
$X=(X_1,\ldots,X_p)\sim\calN(\mu,\Sigma)$ has a p-variate normal
distribution with mean $\mu$ and covariance $\Sigma$. In order to simplify
the notation we set without loss of generality $\mu=0$: for $\mu\neq 0$,
some of the formulae involve the parameter $\mu$ and an intercept column of
$(1,\ldots,1)$ in the design matrices but conceptually, we can proceed as
for the case with $\mu=0$. We
then write $\X=(\X_{\mathrm{obs}},\X_{\mathrm{mis}})$, where $\X$ represents an i.i.d. random sample
of size $n$, $\X_{\mathrm{obs}}$ denotes the set of observed values, and $\X_{\mathrm{mis}}$
the missing data.
\paragraph{Missingness Patterns and Different Parametrizations}
For our purpose it will be convenient to group rows of the matrix $\X$ according to their missingness
patterns \citep{Schafer}. 
We index the unique
missingness patterns that actually appear in our data by $k=1,\ldots,s$. Furthermore, with $o_k\subset \{1,\ldots,p\}$ and
$m_k=\{1,\ldots,p\}\setminus o_k$ we denote the set of observed
variables and the set of missing variables, respectively. $\mathcal{I}_k$ is
the index set of the samples
(row numbers) which belong to pattern $k$, whereas
$\mathcal{I}^c_k=\{1,\ldots,n\}\setminus \mathcal{I}_k$ stands
for the row numbers which do not belong to that pattern. By convention,
samples with all variables observed do not belong to a missingness pattern.

Consider a partition $X=(X_{o_k},X_{m_k})$ of a single Gaussian random
vector. It is well known that
$X_{m_k}|X_{o_k}$ follows a linear regression on $X_{o_k}$ with regression
coefficients $B_{m_k|o_k}$ and covariance $\Sigma_{m_k|o_k}$ given by
\begin{eqnarray}\label{eq:transf}
B_{m_k|o_k}&=&\Sigma_{m_k,o_k}\Sigma_{o_k}^{-1},\\
\Sigma_{m_k|o_k}&=&\Sigma_{m_k}-\Sigma_{m_k,o_k}\Sigma_{o_k}^{-1}\Sigma_{o_k,m_k}.\nonumber
\end{eqnarray}
Consequently, we can write the density $p(x;\Sigma)$ of $X$ as
\begin{eqnarray*}
&&p(x;\Sigma)=p(x_{m_k}|x_{o_k};B_{m_k|o_k},\Sigma_{m_k|o_k})p(x_{o_k};\Sigma_{o_k}),
\end{eqnarray*}
i.e., the density can be characterized by either the parameter $\Sigma$ or
$(\Sigma_{o_k},B_{m_k|o_k},\Sigma_{m_k|o_k})$. With the transformation
(\ref{eq:transf}) we can switch between both parametrizations.

\paragraph{Observed Log-Likelihood and Maximum Likelihood Estimation (MLE)}
A systematic approach to estimate the parameter of interest $\Sigma$ from $\X_{\mathrm{obs}}$ maximizes the observed log-likelihood $\ell(\Sigma;\X_{\mathrm{obs}})$ given by
\begin{eqnarray}\label{eq:observedloglik}
&&\ell(\Sigma;\X_{\mathrm{obs}})=\sum_{i\notin \bigcup_k\mathcal{I}_k}\log p(x_{i};\Sigma)+\sum_{k=1}^{s}\sum_{i\in\mathcal{I}_k}\log p(x_{i, o_k};\Sigma_{o_k}).
\end{eqnarray}

Inference for $\Sigma$ can be based on the observed log-likelihood
(\ref{eq:observedloglik}) if the underlying missing data mechanism is
\emph{ignorable}. The missing data mechanism is said to be \emph{ignorable}
if the probability that an observation is missing may depend on $\X_{\mathrm{obs}}$
but not on $\X_{\mathrm{mis}}$ (\emph{Missing at Random}) and if the parameters of
the data model and the parameters of the missingness mechanism are
\emph{distinct}. For a precise definition see \cite{LittleRubin}.

Explicit maximization of $\ell(\Sigma;\X_{\mathrm{obs}})$ is only possible for
special missing data patterns. Most prominent are examples with a so-called monotone
missing data pattern \citep{LittleRubin,Schafer}, where
$X_1$ is more observed than $X_2$, which is more observed than $X_3$, and
so on. In this case, the observed log-likelihood factorizes and explicit
maximization is achieved by performing several regressions. For a general pattern of missing data, the standard EM
algorithm is often used for optimization of (\ref{eq:observedloglik}). See \cite{Schafer} for a detailed description of the
algorithm. In the next section we present an alternative method for maximizing
the observed log-likelihood. We will argue that this new algorithm is computationally
more efficient than the standard EM.

\section{Pattern Alternating Missing Data Estimation and \\$\ell_1$-Regularization}
For each missingness pattern, indexed by $k=1,\ldots,s$, we introduce some further notation:
\begin{eqnarray*}
&&\X^k=(x_{i,j})\quad \textrm{with}\quad i \in \mathcal{I}_k \quad  j=1,\ldots,p\\
&&\X^{-k}=(x_{i,j})\quad \textrm{with}\quad i \in \mathcal{I}_k^{c} \quad  j=1,\ldots,p.
\end{eqnarray*}
 Thus, $\X^k$ is the $|\mathcal{I}_k|\times p$ submatrix of $\X$ with rows belonging to the $k$th pattern. Similarly, $\X^{-k}$ is the $|\mathcal{I}_k^{c}|\times p$ matrix with rows not belonging to the $k$th pattern. In the same way we define  $\X^{k}_{o_k}, \X^{k}_{m_k}, \X^{-k}_{o_k}$ and $\X^{-k}_{m_k}$. For example,  $\X^{k}_{o_k}$ is defined as the $|\mathcal{I}_k|\times |o_k|$ matrix with
\[\X^k_{o_k}=(x_{i,j})\quad \textrm{with}\quad i \in \mathcal{I}_k, \quad  j \in o_k.\]

\subsection{MLE for Data with a Single Missingness Pattern}\label{sec:singlepattern}
Assume that the data matrix $\X$ has only one single missingness pattern,
denoted by $s$. This is the most simple example of a monotone pattern. The observed log-likelihood factorizes according to: 

\begin{eqnarray}\label{eq:singlepatternloglik}
&&\ell(\Sigma;\X_{\mathrm{obs}})=\sum_{i\in\mathcal{I}_s} \log
p(x_{i,o_s};\Sigma_{o_s})+\sum_{i\in\mathcal{I}_{s}^{c}} \log
p(x_{i};\Sigma)\nonumber\\
&&=\sum_{i=1}^{n} \log
p(x_{i,o_s};\Sigma_{o_s})+\sum_{i\in
  \mathcal{I}_s^{c}}\log p(x_{i,m_s}|x_{i,o_s};B_{m_s|o_s},\Sigma_{m_s|o_s}).
\end{eqnarray}

The left and right part in Equation (\ref{eq:singlepatternloglik}) can be maximized separately. The first part is maximized by the sample covariance of the observed variables based on \emph{all samples}, whereas the second part is maximized by a regression of the missing against observed variables based on only the \emph{fully observed samples}. In formulae:

\begin{eqnarray}\label{eq:obscovsinglemle}
&&\hat{\Sigma}_{o_s}={}^t\X_{o_s}\X_{o_s}/n,
\end{eqnarray}
and
\begin{eqnarray}
&&\hat{B}_{m_s|o_s}={}^t\X^{-s}_{m_s}\X^{-s}_{o_s} ({}^t\X^{-s}_{o_s} \X^{-s}_{o_s})^{-1},\nonumber\\
&&\hat{\Sigma}_{m_s|o_s}={}^t\!(\X^{-s}_{m_s}-\X^{-s}_{o_s}\,{}^t\!\hat{B}_{m_s|o_s})(\X^{-s}_{m_s}-\X^{-s}_{o_s}\,{}^t\!\hat{B}_{m_s|o_s})/|\mathcal{I}_s^{c}|.\label{eq:residcovsinglemle}
\end{eqnarray}

Having these estimates at hand, it is easy to impute the missing data:

\[\hat{x}_{i,m_s}=\hat{B}_{m_s|o_s}{}^t\!x_{i,o_s}\;\textrm{for
  all}\;i\in\mathcal{I}_s,\quad\textrm{or, in matrix notation,}\quad \hat{\X}^{s}_{m_s}=\X^{s}_{o_s}\,{}^t\!\hat{B}_{m_s|o_s}.\]

It is important to note, that, if interested in imputation, only the
regression part of the MLE is needed and the estimate $\hat{\Sigma}_{o_s}$
in (\ref{eq:obscovsinglemle}) is superfluous.
\subsection{MLE for General Missing Data Pattern}\label{sec:patternwise}
We turn now to the general case, where we have more than one missingness
pattern, indexed by $k=1,\ldots,s$. The general idea of the new algorithm is as follows. Assume we have some initial imputations for all missing
values. Our goal is to improve on these imputations. For this purpose, we iterate as follows: 
\begin{itemize}
\item Keep all imputations except those of the $1$st missingness pattern fixed and compute the single
pattern MLE (for the first pattern) as explained in
Section~\ref{sec:singlepattern}. In particular, compute the regression
coefficients of the missing $1$st pattern against all other variables
(treated as ``observed'') based on all samples which do not belong to the
$1$st pattern. 
\item Use the resulting estimates (regression coefficients) to impute
  the missing values from only the $1$st pattern.
\end{itemize} 
Next, turn to the $2$nd pattern and repeat the above steps. In this way we
continue cycling through the different patterns until convergence. 

We now describe the Missingness Pattern Alternating maximization
algorithm (MissPA) which makes
the above idea precise. Let $T={}^t\X\X$ be the sufficient statistic in the
multivariate normal model. Furthermore, denote by $T^{k}={}^t\!(\X^{k})
\X^k$, $T^{-k}={}^t\!(\X^{-k})\X^{-k}=\sum_{l\neq k} T^{l}$. Let $\calT$ and
$\calT^{k}\, (k=1,\ldots,s)$ be some initial guess of $T$ and $T^{k}\,
(k=1,\ldots,s)$, for example, using zero imputation. Our algorithm
proceeds as follows.

\begin{tabular}{lp{13cm}}
  \multicolumn{2}{c}{\rule{0.99\textwidth}{0.95pt}} \\[-0.3ex]
  \multicolumn{2}{l}{\textbf{Algorithm 1: MissPA}} \\[-1.9ex] 
  \multicolumn{2}{c}{\rule{0.99\textwidth}{0.95pt}} \\
  (1) & $\calT$, $\calT^{k}$: initial guess of $T$ and $T^{k}$ $(k=1,\ldots,s)$.\\
  (2) & For $k=1,\ldots,s$ do:\\
  & \hspace{0.2cm} \textbf{M-Step:} Compute the MLE $\hat{B}_{m_k|o_k},$
  and $\hat{\Sigma}_{m_k|o_k}$, based on $\calT^{-k}=\calT-\calT^{k}$: \vspace{0.2cm} \\
       & \hspace{0.8cm} $\hat{B}_{m_k|o_k}=\calT^{-k}_{m_k,o_k}(\calT^{-k}_{o_k,o_k})^{-1},$\\  
       & \hspace{0.8cm} $\hat{\Sigma}_{m_k|o_k}=\left(\calT^{-k}_{m_k,m_k}-\calT^{-k}_{m_k,o_k}(\calT^{-k}_{o_k,o_k})^{-1}\calT^{-k}_{o_k,m_k}\right)/|\mathcal{I}_k^{c}|.$\vspace{0.2cm} \\ 
       & \hspace{0.3cm} \textbf{Partial E-Step:} \vspace{0.2cm} \\
       & \hspace{0.8cm} Set $\calT^{l}=\calT^{l}$ for all $l\neq k$ (this takes
       no time),\vspace{0.1cm} \\
       & \hspace{0.8cm} Set $\calT^{k}=\E[T^{k}|\X^{k}_{o_k},\hat{B}_{m_k|o_k},
       \hat{\Sigma}_{m_k|o_k}],$\vspace{0.1cm} \\
       & \hspace{0.8cm} Update $\calT=\calT^{-k}+\calT^k$.\vspace{0.2cm}\\
  (3) & Repeat step (2) until some convergence criterion is met.\vspace{0.2cm}\\
  (4)& Compute the final maximum likelihood estimator $\hat{\Sigma}$ via:\vspace{0.2cm}\\
  &\hspace{0.3cm}$\hat{\Sigma}_{o_s}=\calT_{o_s,o_s}/n$,
  $\hat{\Sigma}_{m_s,o_s}=\hat{B}_{m_s|o_s}\hat{\Sigma}_{o_s}$ and $\hat{\Sigma}_{m_s}=\hat{\Sigma}_{m_s|o_s}+\hat{B}_{m_s|o_s}\hat{\Sigma}_{o_s,m_s}$.\\[-1.0ex]
  \multicolumn{2}{c}{\rule{0.99\textwidth}{0.95pt}}
\end{tabular}
%
%

Note, that we refer to the maximization step as M-Step and to the imputation step
as \emph{partial} E-Step. The word partial refers to the fact that the expectation
is only performed with respect to samples belonging to the current pattern. The partial E-Step of our algorithm takes the following simple form:
\begin{eqnarray*}
&&\calT^{k}_{o_k,m_k}={}^t\!(\X_{o_k}^{k})\hat{\X}_{m_k}^{k},\\
&&\calT^{k}_{m_k,m_k}={}^t\!(\hat{\X}_{m_k}^{k})\hat{\X}_{m_k}^k+|\mathcal{I}_k|\hat{\Sigma}_{m_k|o_k},
\end{eqnarray*}
where $\hat{\X}^{k}_{m_k}=\E[\X^{k}_{m_k}|\X^{k}_{o_k},\hat{B}_{m_k|o_k},\hat{\Sigma}_{m_k|o_k}]=\X^{k}_{o_k}{}^t\!\hat{B}_{m_k|o_k}$.

Our algorithm does not require an evaluation of $\hat{\Sigma}_{o_k}$ in the
M-Step, as it is not used in the following partial E-Step. But, if
we are interested in the observed log-likelihood or the maximum likelihood
estimator $\hat{\Sigma}$ at convergence, we compute $\hat{\Sigma}_{o_s}$
(at convergence), use it together with $\hat{B}_{m_s|o_s}$ and
$\hat{\Sigma}_{m_s|o_s}$ to get $\hat{\Sigma}$ via the transformations
(\ref{eq:transf}) as explained in step~(4). 

MissPA is computationally more efficient than the standard EM for missing data: one cycle through all
patterns ($k=1,\ldots,s$) takes about the same time as one iteration
of the standard EM. But our algorithm makes more progress since the information
from the partial E-Step is utilized immediately to perform the next
M-Step. We will demonstrate empirically the gain of computational
efficiency in Section~\ref{sec:efficiency}. The new MissPA generalizes the standard EM in two ways. Firstly, MissPA alternates between different complete data spaces in the
  sense of \cite{sage1994}. Secondly, the E-Step
  is performed incrementally \citep{incremental}. In Section~\ref{sec:theory} we will
  expand on these generalizations and we will provide an appropriate framework
which allows analyzing the convergence properties of MissPA. 

Finally, a small modification of MissPA, namely replacing in Algorithm~1  the M-Step by

 \hspace{1cm}  \vspace{0.2cm}\textbf{M-Step2:} Compute the MLE $\hat{B}_{m_k|o_k},$
 and $\hat{\Sigma}_{m_k|o_k}$, based on $\calT$: \\\vspace{0.2cm}
\hspace{1.6cm}$\hat{B}_{m_k|o_k}=\calT_{m_k,o_k}(\calT_{o_k,o_k})^{-1}$\\
\vspace{0.2cm}
\hspace{1.6cm}$\hat{\Sigma}_{m_k|o_k}=\left(\calT_{m_k,m_k}-\calT_{m_k,o_k}(\calT_{o_k,o_k})^{-1}\calT_{o_k,m_k}\right)/n,$

results in an alternative algorithm. We show in
Section~\ref{sec:theory} that Algorithm~1 with M-Step2 is equivalent to an incremental EM in the sense of
\cite{incremental}.

\subsection{High-Dimensionality and Lasso
  Penalty}\label{sec:patternwiselasso}
The M-Step of Algorithm~1 is basically a multivariate regression of
the missing ($X_{m_k}$) against the observed variables ($X_{o_k}$). In
a high-dimensional framework with $p\gg n$ the number of observed
variables $|o_k|$ will be large and therefore some
regularization is necessary. The main idea is, in order to regularize,
to replace the regressions with a Lasso \citep{tibshirani96regression}. We give now the details.

\paragraph{Estimation of $B_{m_k|o_k}$:}
The estimation of the  multivariate regression coefficients in the
M-Step2 can be expressed as $|m_k|$ separate minimization problems of the form
\begin{eqnarray*}
&&\hat{B}_{j|o_k}=\argmin_{\beta}-\calT_{j,o_k}\beta+{}^t\!\beta\calT_{o_k,o_k}\beta/2,
\end{eqnarray*}
where $j\in m_k$. Here, $\hat{B}_{j|o_k}$ denotes the $j$th row vector of the
$(|m_k|\times|o_k|)$-matrix $\hat{B}_{m_k|o_k}$ and represents the regression of
variable $j$ against the variables from $o_k$.

Consider now the objective function
\begin{eqnarray}\label{sec:lassoobjective}
&&-\calT_{j,o_k}\beta+{}^t\!\beta\calT_{o_k,o_k}\beta/2+\lambda\|\beta\|_1,
\end{eqnarray}
with an additional Lasso penalty. Instead of minimizing (\ref{sec:lassoobjective}) with respect to $\beta$
(for all $j\in m_k$), it is computationally much
more efficient to improve it coordinate-wise only
from the old parameters (computed in the last cycle through all
patterns). For that purpose, let $B_{m_k|o_k}^{(r)}$ be the regression coefficients
for pattern $k$ in cycle $r$ and $B_{j|o_k}^{(r)}$ its $j$th row vector. In cycle $r+1$
we compute $B_{j|o_k}^{(r+1)}$ by
minimizing (\ref{sec:lassoobjective}) with respect to each of the
components of $\beta$, holding the other components fixed at their current value.
Closed-form updates have the form:
\begin{eqnarray}\label{eq:thresholding}&&B_{j|l}^{(r+1)} =
\frac{\mathop{Soft}\big(\calT_{l,l}B_{j|l}^{(r)}-S_l,\lambda\big)}
{\calT_{l,l}},\qquad \textrm{for all $l\in o_k$,}
\end{eqnarray}
where
\begin{itemize}
\item $B_{j|l}^{(r+1)}$ is the $l$th component of $B_{j|o_k}^{(r+1)}$
  equal to the element $(j,l)$ of matrix $B_{m_k|o_k}^{(r+1)}$.
\item
$S_l$, the gradient of
$-\calT_{j,o_k}\beta+{}^t\!\beta\calT_{o_k,o_k}\beta/2$ with respect to
$\beta_l$, which equals
\begin{eqnarray}\label{eq:gradient}S_l=-\calT_{j,l}+\sum_{\substack{v<l
      \\v\in
      o_k}}\calT_{l,v}B_{j|v}^{(r+1)}+\calT_{l,l}B_{j|l}^{(r)}
+\sum_{\substack{v>l\\v\in o_k}}\calT_{l,v}B_{j|v}^{(r)}.
\end{eqnarray}
\item
  $\mathop{Soft}(z,\lambda)=\left\{\begin{array}{ll}z-\lambda&\textrm{if}\;z>\lambda
      \\ z+\lambda&\textrm{if}\;z< -\lambda \\0&\textrm{if}\;\abs{z}\leq
      \lambda \end{array}\right.$, is the standard
  soft-thresholding operator.
\end{itemize}

In a sparse setup the soft-thresholding update (\ref{eq:thresholding}) can be
evaluated very quickly as $l$ varies and often coefficients which are zero
remain zero after thresholding. See also the \emph{naive}- or
\emph{covariance} \emph{update} idea of \cite{friedmanetal08} for efficient
computation of (\ref{eq:thresholding}) and (\ref{eq:gradient}).
\paragraph{Estimation of $\Sigma_{m_k|o_k}$:}
We update the residual covariance matrix as:
\begin{eqnarray}\label{eq:residualcov}\Sigma_{m_k|o_k}^{(r+1)}=\Big(\calT_{m_k,m_k}-\calT_{m_k,o_k}{}^t\!B_{m_k|o_k}^{(r+1)}-B_{m_k|o_k}^{(r+1)}\calT_{o_k,m_k}+
  B_{m_k|o_k}^{(r+1)}\calT_{o_k,o_k}{}^t\!B_{m_k|o_k}^{(r+1)}\Big)/n.
\end{eqnarray}
Formula (\ref{eq:residualcov}) can be viewed as a generalized version of
Equation (\ref{eq:residcovsinglemle}), when multiplying out the matrix
product in (\ref{eq:residcovsinglemle}) and taking expectations.

Our regularized algorithm, the MissPALasso, is summarized in
Table~\ref{tab:alg2}. Note, that we update the sufficient
statistic in the partial E-Step according to
$\calT=\gamma\calT+\calT^k$ where $\gamma=1-|\mathcal{I}_k|/n$. This
update, motivated by \cite{nowlan91}, calculates $\calT$ as an
exponentially decaying average of recently-visited data points. It prevents MissPALasso from storing $\calT^k$ for all
$k=1,\ldots,s$ which gets especially cumbersome for large
$p$'s. 
As we are mainly interested in estimating the missing values, we will output the
data matrix with missing values imputed by the regression
coefficients $\hat{B}_{m_k|o_k}$ ($k=1,\ldots,s$) as indicated in step (4)
of Table~\ref{tab:alg2}. MissPALasso provides not only the imputed data
matrix $\hat{\X}$ but also $\hat{\calT}$, the completed version of the
sufficient statistic ${}^t\!\X\X$. The latter can be very useful if MissPALasso
is used as a pre-processing step followed by a learning method which is
expressible in terms of the sufficient statistic. Examples include
regularized regression (e.g., Lasso), discriminant analysis, or estimation of
directed acyclic graphs with the PC-algorithm \citep{spirtes}.

By construction, the regression estimates $\hat{B}_{m_k|o_k}$ are sparse,
due to the employed $\ell_1$-penalty, and therefore, imputation of missing
values $\hat{\X}^{k}_{m_k}=\X^{k}_{o_k}\,{}^t\!\hat{B}_{m_k|o_k}$ is based
on sparse regressions. This is in sharp contrast to the MissGLasso approach
(see Section~\ref{sec:performance}) which places sparsity on
$\Sigma^{-1}$. But this does not imply that regressions of variables in
$m_k$ on variables in $o_k$ are sparse since the inverse of sub-matrices of
a sparse $\Sigma^{-1}$ are not sparse in general. MissPALasso employs
another type of sparsity and this seems to be the main reason for its
better statistical performance than MissGLasso.
\\
\begin{remark}
In practice, we propose to compute MissPALasso for a decreasing
sequence of values for $\lambda$, using each solution as a warm start for
the next problem with smaller $\lambda$. This pathwise strategy is
computationally very attractive and our algorithm
converges (for each $\lambda$) after a few cycles.
\end{remark}
\begin{table}[!h]
  \begin{tabular}{lp{15cm}}
    \multicolumn{2}{l}{\rule{0.995\textwidth}{0.95pt}} \\[-0.3ex]
    \multicolumn{2}{l}{\hspace{0.2cm}\textbf{Algorithm 2: MissPALasso}} \\[-1.9ex] 
    \multicolumn{2}{l}{\rule{0.995\textwidth}{0.95pt}} \\
    (1) &Set $r=0$ and start with initial guess for $\calT$ and $B_{m_k|o_k}^{(0)}$ ($k=1,\ldots,s$).\vspace{0.2cm} \\
    (2)&In cycle $r+1$; for $k=1,\ldots,s$ do:\vspace{0.1cm}\\
        &\hspace{0.3cm}\textbf{Penalized M-Step2:} \vspace{0.1cm}\\
&\hspace{0.55cm}For all $j \in m_k$, compute $B_{j|o_k}^{(r+1)}$
by improving $-\calT_{j,o_k}\beta+{}^t\!\beta\calT_{o_k,o_k}\beta/2+\lambda\|\beta\|_1$\vspace{0.00cm}\\
&\hspace{0.55cm}in a coordinate-wise manner from $B_{j|o_k}^{(r)}$.\vspace{0.3cm}\\
&\hspace{0.55cm}Set
$\Sigma_{m_k|o_k}^{(r+1)}\!=\Big(\calT_{m_k,m_k}\!-\calT_{m_k,o_k}{}^t\!B_{m_k|o_k}^{(r+1)}\!-B_{m_k|o_k}^{(r+1)}\calT_{o_k,m_k}\!+ B_{m_k|o_k}^{(r+1)}\calT_{o_k,o_k}{}^t\!B_{m_k|o_k}^{(r+1)}\Big)/n.$\vspace{0.1cm}\\

        & \hspace{0.3cm}\textbf{Partial E-Step:}\vspace{0.1cm} \\
        & \hspace{0.55cm}Set $\calT^{k}=\E[T^{k}|\X^k_{o_k},B_{m_k|o_k}^{(r+1)},
  \Sigma^{(r+1)}_{m_k|o_k}],$\vspace{0.1cm} \\
       & \hspace{0.55cm}Update $\calT=\gamma\calT+\calT^k$ where $\gamma=1-|\mathcal{I}_k|/n$.\vspace{0.1cm}\\
 &Increase: $r\leftarrow r+1.$ \vspace{0.25cm}\\
    (3)&Repeat step (2) until some convergence criterion is met.\vspace{0.2cm}\\
    (4)&Output the imputed data matrix $\hat{\X}$, with missing values
    estimated by:\vspace{0.1cm}\\
    &$\hat{\X}^{k}_{m_k}=\X^{k}_{o_k}\,{}^t\!\hat{B}_{m_k|o_k}$, $k=1,\ldots,s$.\vspace{0.2cm}\\[-1.0ex]
          \multicolumn{2}{l}{\rule{0.995\textwidth}{0.95pt}}
  \end{tabular}
\caption{MissPALasso. In the $k$th M-Step of cycle $r+1$, instead of a
  multivariate Lasso regression, a coordinate descent approximation of
  the corresponding Lasso problem is performed. Regression
  coefficients $B^{(r)}_{m_k|o_k}$, $k=1,\ldots,s$, are stored in
  sparse matrix format.}
\label{tab:alg2}
\end{table}

\section{Numerical Experiments}
\subsection{Performance of MissPALasso}\label{sec:performance}
In this section we will explore the performance of MissPALasso
developed in Section~\ref{sec:patternwiselasso}. We compare our new method
with alternative ways of imputing missing values in high-dimensional
data. We consider the following methods:
\begin{itemize}
\item \emph{KnnImpute:} Impute the missing values by the K-nearest
  neighbors imputation method introduced by
  \cite{knnimpute2001}.
\item \emph{SoftImpute:} The soft imputation algorithm is proposed by
  \cite{softimpute2008} in order to solve the matrix completion
  problem. They propose to approximate the incomplete data matrix $\X$ by a
  complete (low-rank) matrix $\mathbf{Z}$ minimizing
\[\frac{1}{2} \sum_{(i,j)\in \Omega}(z_{ij}-x_{ij})^2 + \lambda
\|\mathbf{Z}\|_*.\]
Here, $\Omega$ denotes the indices of observed entries and
$\|\mathbf{Z}\|_*$ is the nuclear norm, or the sum of the singular
values. The missing values of $\X$ are imputed by the corresponding values of $\mathbf{Z}$.
\item \emph{MissGLasso:} Compute $\hat{\Sigma}$ by
  minimizing $-\ell(\Sigma;\X_{\mathrm{obs}})+\lambda\|\Sigma^{-1}\|_1,$ where
  $\|\Sigma^{-1}\|_1$ is the entrywise $\ell_1$-norm. Then, use this
  estimate to impute the missing values by conditional mean imputation. \emph{MissGLasso} is described in \cite{missglasso2010}.
\item \emph{MissPALasso:} This is the method introduced in Section~\ref{sec:patternwiselasso}.
\end{itemize}
To assess the performances of the methods we use the normalized root mean
squared error \citep{bpcaimpute} which is defined by
\begin{eqnarray*}
\textrm{NRMSE}&=&\sqrt{\frac{\textbf{mean}\left((\X^{\mathrm{true}}-\hat{\X})^2\right)}{\textbf{var}\left(\X^{\mathrm{true}}\right)}}.
\end{eqnarray*}
Here, $\X^{\mathrm{true}}$ is the original data matrix (before deleting values) and
$\hat{\X}$ is the imputed matrix. With \textbf{mean} and \textbf{var} we abbreviate the
empirical mean and variance, calculated over only the missing entries.

All methods involve one tuning parameter. In KnnImpute we have
to choose the number $K$ of nearest neighbors, while SoftImpute, MissGLasso
and MissPALasso involve
a regularization parameter which is always denoted by $\lambda$. In all of our experiments we choose
the tuning parameters to obtain optimal performance in terms of NRMSE. 

\subsubsection{Simulation Study}\label{sec:sim}
We consider both high- and a low-dimensional MVN models with
$\sim\calN_p(0,\Sigma)$ where
\begin{itemize}
\item {\bf Model 1:} $p=50$ and $500$;\\$\Sigma$: block diagonal with $p/2$ blocks of
  the form 
{$\bigl(\begin{smallmatrix} 
1&0.9\\ 
0.9&1
\end{smallmatrix}\bigr).$}

\item {\bf Model 2:} $p=100$ and 1000;\\$\Sigma$: two blocks $\textrm{B}_1$,
  $\textrm{B}_2$ each of size $\frac{p}{2}\times\frac{p}{2}$ with $\textrm{B}_1
  =\textrm{I}_{\frac{p}{2}}$ and $(\textrm{B}_2)_{j,j'}=0.9^{|j-j'|}$.

\item {\bf Model 3:} $p=55$ and 496;\\$\Sigma$: block diagonal with
  $b=1,\ldots,10$ for $p=55$ and $b=1,\ldots,31$ for $p=496$ (increasing)
  blocks $\textrm{B}_b$ of the size $b\times b$, with
  $(\textrm{B}_b)_{j,j'}=0.9$ $(j\neq j')$ and $(\textrm{B}_b)_{j,j}=1$.

\item {\bf Model 4:} $p=100$ and 500;\\$\Sigma_{j,j'}=0.9^{|j-j'|}$ for
  $j,j'=1,\ldots,p.$
\end{itemize}  

For all four settings we perform 50 independent simulation runs. In each
run we generate $n=50$ i.i.d. samples from the model. We then delete
randomly $5\%, 10\%$ and $15\%$ of the values in the data matrix, apply an
imputation method and compute the NRMSE. The results of the different
imputation methods (tuning parameters such that NRMSE is minimal) are reported in Table~\ref{tab:sim1}
for the low-dimensional models and Table~\ref{tab:sim2} for the high-dimensional
models. MissPALasso is very competitive in all setups. SoftImpute works
rather poorly, perhaps because the resulting data matrices are not well
approximable by low-rank matrices. KnnImpute works very well in model 1 and
model 4. Model 1, where each variable is highly correlated with its
neighboring variable, represents an example which fits well into the
KnnImpute framework. However, in model 2 and model 3, KnnImpute performs
rather poorly. The reason is that with an inhomogeneous covariance matrix,
as in model 2 and 3, the optimal number of nearest neighbors is varying
among the different blocks, and a single parameter $K$ is too
restrictive. For example in model 2, a variable from the first block
is not correlated to any other variable, 
whereas a variable from the second block is correlated to other
variables. Except for the low-dimensional model 3 MissGLasso is
inferior to MissPALasso. Furthermore, MissPALasso strongly outperforms
MissGLasso with respect to computation time (see Figure~\ref{fig:timing} in Section~\ref{sec:efficiency}). Interestingly, all methods exhibit a quite large
NRMSE in the high-dimensional model 3. They seem to have problems coping
with the complex covariance structure in higher dimensions. If we look at
the same model but with $p=105$ the NRMSE for 5\% missing values is: 0.85 for KnnImpute, 0.86 for SoftImpute, 0.77
for MissGLasso and 0.77 for MissPALasso. This indicates an increase in NRMSE
according to the size of $p$. Arguably, we consider here only multivariate
normal models which are ideal, from a distributional point of view, for
MissGLasso and our MissPALasso. The more interesting case will be with real
data (all from genomics) where model assumptions never hold exactly.
 
\begin{table}[!h]
\begin{center}
\tabcolsep=6.0pt
\begin{tabular}{|lr|c|c|c|c|}
  \hline
 && KnnImpute & SoftImpute & MissGLasso & MissPALasso \\ 
  \hline\hline
Model 1 &5\% &0.4874 (0.0068)&0.7139  (0.0051)&0.5391  (0.0079)&0.5014  (0.0070)\\ 
p=50    &10\% &0.5227 (0.0051)&0.7447  (0.0038)&0.5866  (0.0057)&0.5392  (0.0055)\\ 
        &15\% &0.5577  (0.0052)&0.7813  (0.0037)&0.6316  (0.0048)&0.5761  (0.0047)\\ 
\hline

Model 2 &5\% &0.8395  (0.0101)&0.8301  (0.0076)&0.7960  (0.0082)&0.7786  (0.0075)\\ 
p=100   &10\% &0.8572  (0.0070)&0.8424  (0.0063)&0.8022  (0.0071)&0.7828  (0.0066)\\ 
        &15\% &0.8708  (0.0062)&0.8514  (0.0053)&0.8082  (0.0058)&0.7900  (0.0054)\\ 
\hline
Model 3 &5\% &0.4391  (0.0061)&0.4724 (0.0050)&0.3976  (0.0056)&0.4112  (0.0058)\\ 
p=55    &10\% &0.4543  (0.0057)&0.4856 (0.0042)&0.4069  (0.0047)&0.4155  (0.0047)\\  
  &15\% &0.4624  (0.0054)&0.4986  (0.0036)&0.4131  (0.0043)&0.4182  (0.0044)\\ 
\hline
Model 4 &5\% &0.3505  (0.0037)&0.5515  (0.0039)&0.3829  (0.0035)&0.3666  (0.0031)\\ 
p=100   &10\% &0.3717 (0.0033)&0.5623  (0.0033)&0.3936  (0.0027)&0.3724 (0.0026)\\ 
        &15\% &0.3935  (0.0032)&0.5800  (0.0031)&0.4075  (0.0026)&0.3827  (0.0026)\\ \hline
\end{tabular}
\caption{Average (SE) NRMSE of KnnImpute, SoftImpute, MissGLasso and
  MissPALasso with different degrees of missingness in the low-dimensional
  models.}
\label{tab:sim1}
\end{center}
\end{table}

\begin{table}[!h]
\begin{center}
\tabcolsep=6.0pt
\begin{tabular}{|lr|c|c|c|c|}
  \hline
 && KnnImpute & SoftImpute & MissGLasso & MissPALasso \\ 
  \hline\hline
Model 1 &5\%  &0.4913 (0.0027)&0.9838  (0.0006)&0.6705  (0.0036)&0.5301  (0.0024)\\ 
p=500   &10\% &0.5335 (0.0020)&0.9851  (0.0005)&0.7613  (0.0031)&0.5779  (0.0019)\\ 
        &15\% &0.5681 (0.0016)&0.9870  (0.0004)&0.7781  (0.0013)&0.6200  (0.0015)\\ 
\hline

Model 2 &5\%  &0.8356  (0.0020)&0.9518  (0.0009)&0.8018 (0.0012)&0.7958  (0.0017)\\ 
p=1000  &10\% &0.8376  (0.0016)&0.9537  (0.0007)&0.8061 (0.0002) &0.7990  (0.0013)\\ 
        &15\% &0.8405  (0.0014)&0.9562  (0.0006)& 0.8494 (0.0080) &0.8035  (0.0011)\\ 
\hline
Model 3 &5\%  &1.0018  (0.0009)&0.9943 (0.0005)&0.9722 (0.0013)&0.9663  (0.0010)\\ 
p=496   &10\% &1.0028  (0.0007)&0.9948 (0.0004)&0.9776 (0.0010)&0.9680  (0.0007)\\  
        &15\% &1.0036  (0.0006)&0.9948 (0.0003)&0.9834 (0.0010)&0.9691  (0.0007)\\ 
\hline
Model 4 &5\%  &0.3487  (0.0016)&0.7839  (0.0020)&0.4075 (0.0016)&0.4011  (0.0016)\\ 
p=500   &10\% &0.3721  (0.0014)&0.7929  (0.0015)&0.4211 (0.0012)&0.4139  (0.0013)\\ 
        &15\% &0.3960  (0.0011)&0.8045  (0.0014)&0.4369 (0.0012)&0.4292  (0.0014)\\ \hline
\end{tabular}
\caption{Average (SE) NRMSE of KnnImpute, SoftImpute, MissGLasso and
  MissPALasso with different degrees of missingness in the high-dimensional
  models.}
\label{tab:sim2}
\end{center}
\end{table}


\subsubsection{Real Data Examples}\label{sec:real}
We consider the following four publicly available datasets:

\begin{itemize}
\item \textbf{Isoprenoid gene network in Arabidopsis thaliana:} The number of
  genes in the network is $p=39$. The number of observations (gene
  expression profiles), corresponding to different experimental conditions,
  is $n=118$. More details about the data can be found in \cite{wille}.
\item \textbf{Colon cancer:} In this dataset, expression levels of
  40 tumor and 22 normal colon tissues ($n=62$) for $p=2000$ human genes are
  measured. For more information see \cite{alon99}.
\item \textbf{Lymphoma:} This dataset, presented in \cite{lymphoma2000},
  contains gene expression levels of 42 samples of diffuse large B-cell
  lymphoma, 9 observations of follicular lymphoma, and 11 cases of chronic
  lymphocytic leukemia. The total sample size is $n=62$ and $p=1332$
  complete measured expression profiles are documented.
\item \textbf{Yeast cell-cycle:} The dataset, described in
  \cite{spellman98}, monitors expressions of 6178 genes. The data consists
  of four parts, which are relevant to alpha factor ($18$ samples),
  elutriation ($14$ samples), \emph{cdc15} ($24$ samples), and \emph{cdc28}
  ($17$ samples). The total sample size is $n=73$. We use the $p=573$
  complete profiles in our study.
\end{itemize}

For all datasets we standardize the columns (genes) to zero mean and
variance one. In order to compare the performance of the different
imputation methods we randomly delete values to obtain an
overall missing rate of $5\%$, $10\%$ and $15\%$. Table~\ref{tab:realdata}
shows the results for 50 simulation runs, where in each run another random
set of values is deleted.

\begin{table}[!h]
\begin{center}
\tabcolsep=3.0pt
\begin{tabular}{|lr|c|c|c|c|}
  \hline
 && KnnImpute & SoftImpute & MissGLasso & MissPALasso \\ 
  \hline\hline
Arabidopsis & 5\% &0.7732 (0.0086)&0.7076  (0.0065)&0.7107  (0.0076)&0.7029  (0.0077)\\ 
n=118       &10\% &0.7723 (0.0073)&0.7222  (0.0052)&0.7237  (0.0064)&0.7158  (0.0060)\\ 
p=39        &15\% &0.7918  (0.0050)&0.7369  (0.0041)&0.7415 (0.0053)&0.7337 (0.0050)\\ 
\hline
Colon cancer & 5\% &0.4884  (0.0011)&0.4921  (0.0011)&    -         &0.4490  (0.0011)\\ 
n=62         &10\% &0.4948  (0.0008)&0.4973  (0.0006)&    -         &0.4510  (0.0006)\\ 
p=2000       &15\% &0.5015  (0.0007)&0.5067  (0.0006)&    -         &0.4562  (0.0007)\\ 
\hline
Lymphoma & 5\% &0.7357  (0.0014)&0.6969  (0.0008)&      -         &0.6247  (0.0012)\\ 
n=62     &10\% &0.7418  (0.0009)&0.7100  (0.0006)&      -         &0.6384 (0.0009)\\ 
p=1332   &15\% &0.7480  (0.0007)&0.7192  (0.0005)&      -         &0.6525 (0.0008)\\ 
\hline
Yeast cell-cycle & 5\% &0.8083 (0.0018)&0.6969 (0.0012)& - & 0.6582 (0.0016)\\ 
n=73             &10\% &0.8156 (0.0011)&0.7265 (0.0010)& - & 0.7057 (0.0013)\\ 
p=573            &15\% &0.8240 (0.0009)&0.7488 (0.0007)& - & 0.7499 (0.0011)\\ \hline
\end{tabular}
\caption{Average (SE) NRMSE of KnnImpute, SoftImpute, MissGLasso and
  MissPALasso for different real datasets from genomics. The \textsf{R}
  implementation of MissGLasso is not able to handle real datasets of such
  high dimensionality.}
\label{tab:realdata}
\end{center}
\end{table}

MissPALasso exhibits in all setups the lowest averaged
NRMSE. MissGLasso performs nearly as well as MissPALasso on
the Arabidopsis data. However, its \textsf{R} implementation cannot cope
with large values of $p$. If we would restrict our analysis to the 100
variables exhibiting the most variance we would see that MissGLasso
performs slightly less than MissPALasso (results not included). Compared to
KnnImpute, SoftImpute works well for all datasets. Interestingly, KnnImpute
performs for all datasets much inferior to MissPALasso. In light of the
simulation results of Section~\ref{sec:sim}, a reason for the poor
performance could be that KnnImpute has difficulties with the inhomogeneous
correlation structure between different genes which is plausible to be
present in real datasets.

\begin{figure}[p]
\begin{centering}
  \includegraphics[scale=0.85]{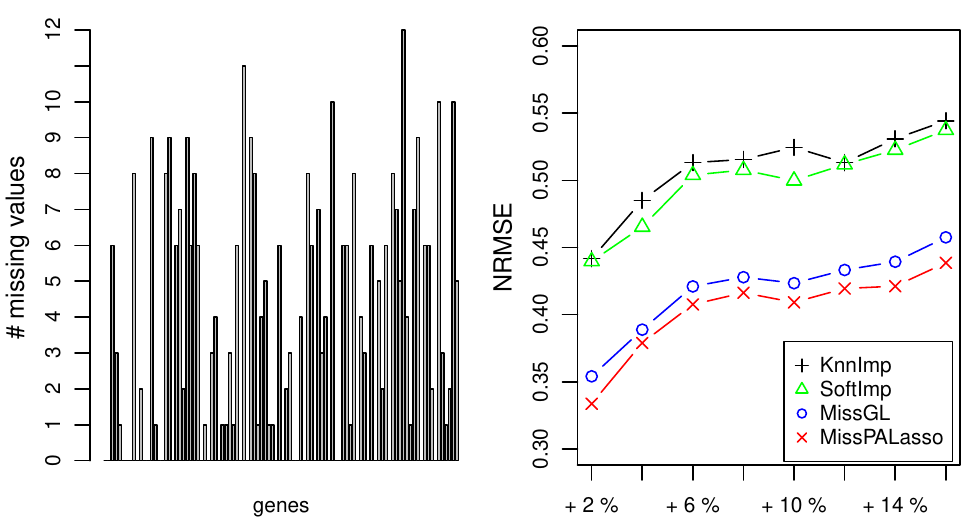} 

  \caption[]
  {Lymphoma dataset. Left panel: Barplots which count the number of missing
    values for each of the 100 genes. Right panel: NRMSE for KnnImpute,
    SoftImpute, MissGLasso and MissPALasso if we introduce additional $2\%,
    4\%, 6\% ,\ldots,16\%$ missings.}
\label{fig:lymphoma}
\end{centering}
\end{figure}

\begin{figure}[p]
\begin{centering}
  \includegraphics[scale=0.85]{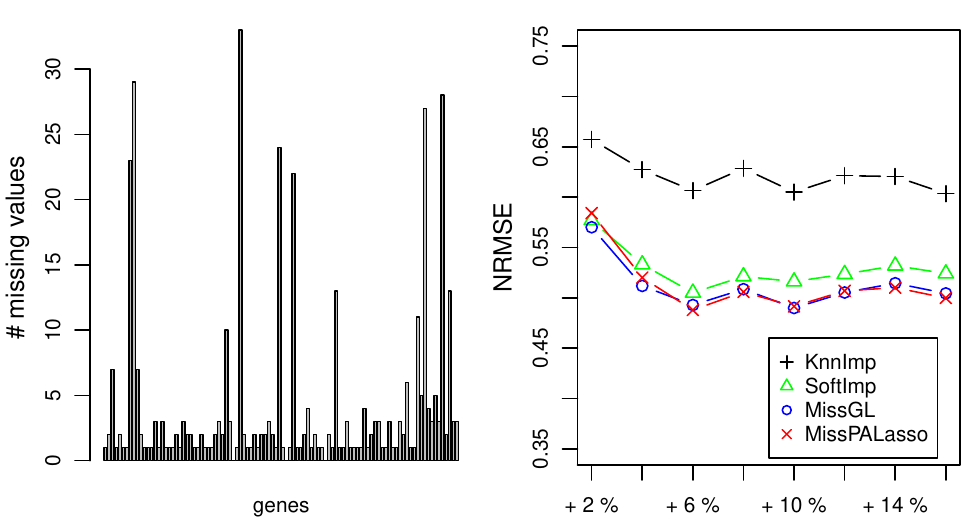} 

  \caption[]
  {Yeast cell-cycle dataset. Left panel: Barplots which count the number of
    missing values for each of the 100 genes. Right panel: NRMSE for
    KnnImpute, SoftImpute, MissGLasso and MissPALasso if we introduce
    additional $2\%, 4\%, 6\% ,\ldots,16\%$
    missings.}
\label{fig:yeastcellcycle}
\end{centering}
\end{figure}

To investigate the effect of already missing values on the imputation
performance of the compared methods we use the original lymphoma and yeast
cell-cycle datasets which already have ``real'' missing values. We only
consider the 100 most variable genes in these datasets to be able to
compare all four methods with each other. From the left panel of
Figures~\ref{fig:lymphoma} and \ref{fig:yeastcellcycle} we can read off how
many values are missing for each of the 100 variables. In the right panel
of Figures~\ref{fig:lymphoma} and \ref{fig:yeastcellcycle} we show how well
the different methods are able to estimate $2 \%, 4\%, 6\%\ldots, 16\%$ of
additionally deleted entries. 

\subsection{Computational Efficiency}\label{sec:efficiency}
\begin{figure}[htbp]
\begin{centering}
  \includegraphics[scale=0.98]{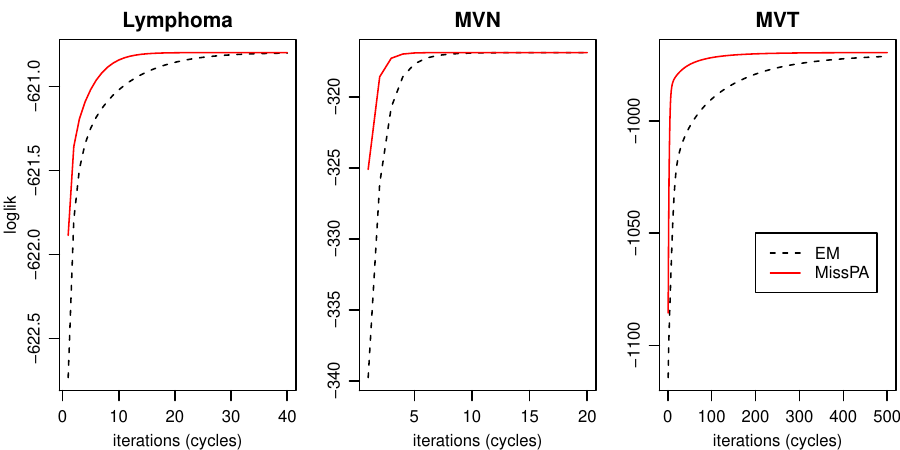} 
  \caption[]
{Log-likelihood as a function of the number of iterations (cycles) for the
  standard EM and MissPA. Left panel: subset of the lymphoma data
  ($n=62$, $p=10$ and $22.85\%$ missing values). Middle panel: random
  sample of the size $62\times
10 $ from the multivariate normal model with the same missing entries as in the reduced lymphoma data. Right panel: random
sample of the size $62\times 10 $ from the multivariate t-model again with
the same missing values.}
\label{fig:loglikvsiter}
\end{centering}
\end{figure}

We first compare the computational efficiency of MissPA with the standard EM described for example in
\cite{Schafer}. The reason why our algorithm takes less time to converge is
because of the more frequent updating of the latent distribution in the
M-Steps. A key attribute of MissPA is that the computational cost of
one cycle through all patterns is the same as the cost of a single E-Step
of the standard EM which requires computation of the regression
parameters from $\hat\Sigma$ obtained in the previous M-Step. This is a big contrast to the incremental EM, mostly
applied to finite mixtures \citep{Thiesson2001,McLachlanNg2003}, where
there is a trade-off between the additional computation time per cycle, or
``scan'' in the language of \cite{McLachlanNg2003}, and the fewer number of
``scans'' required because of the more frequent updating after each partial
E-Step. The speed of convergence of the standard EM and MissPA for
three datasets are shown in Figure~\ref{fig:loglikvsiter}, in which the
log-likelihood is plotted as a function of the number of iterations
(cycles). The left panel corresponds to the subset of the lymphoma dataset
when only the ten genes with highest missing rate are used. This results in
a $62\times 10 $ data matrix with $22.85 \% $ missing values. For the
middle panel we draw a random sample of size $62\times 10 $ from
$\calN_{10}(0,\Sigma)$, $\Sigma_{j,j'}=0.9^{|j-j'|}$, and delete the same
entries which are missing in the reduced lymphoma data. For the right panel
we draw from the multivariate t-model with degrees of freedom equal to one
and again with the same values deleted. As can be seen, MissPA
converges after fewer cycles. A very extreme example is obtained with the
multivariate t-model where the standard EM reaches the log-likelihood level
of MissPA about 400 iterations later. We note here, that the result in
the right panel highly depends on the realized random sample. With other
realizations, we get less and more extreme results than the one shown in
Figure~\ref{fig:loglikvsiter}.

\begin{figure}[htbp]
\begin{centering}
  \includegraphics[scale=0.5]{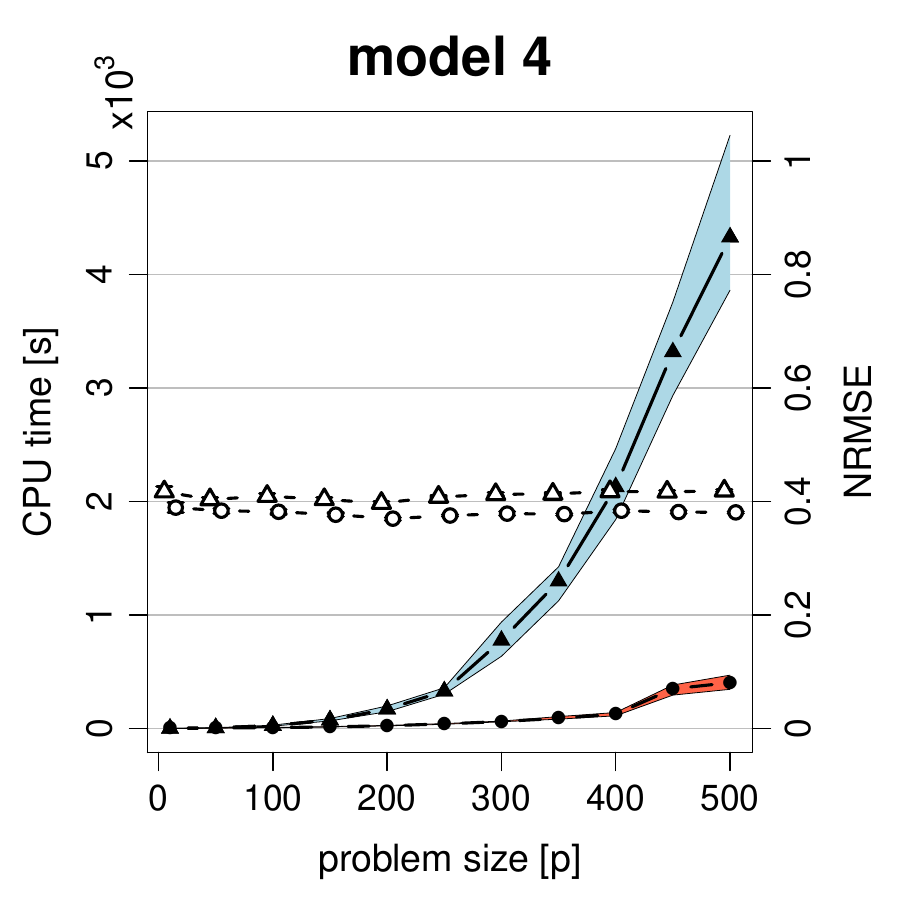}\includegraphics[scale=0.5]{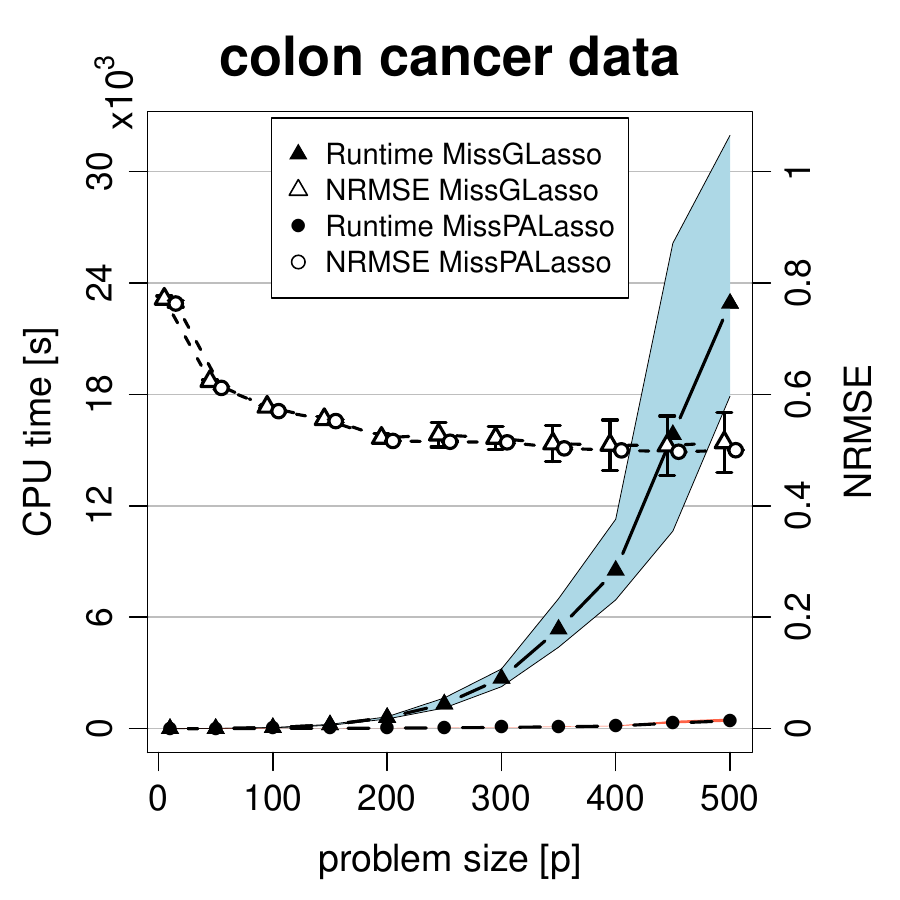}  
  \caption[]
  {CPU times (filled points, left axis) and NRMSE (hollow points, right
    axis) vs. problem size $p$ of MissPALasso (circles) and MissGLasso (triangles) in
    simulation model 4 (left panel) and the colon cancer data (right
    panel). MissPALasso and MissGLasso are applied on a grid
    of thirty $\lambda$ values. The shaded area shows the full range
    of CPU times over 10 simulation runs. Measurements of NRMSE include
    standard error bars which are due to their small size ($\sim10^{-3}$)
    mostly not visible except for MissGLasso in the real data example.
  }
\label{fig:timing}
\end{centering}
\end{figure}

We end this section by illustrating the computational timings of MissPALasso and MissGLasso implemented with the statistical computing
language \textsf{R}. We consider two settings. Firstly, model 4 of
Section~\ref{sec:sim} with $n=50$ and a growing number of variables $p$
ranging from 10 to 500. Secondly, the colon cancer dataset from
Section~\ref{sec:real} with $n=62$ and also a growing number of variables
where we sorted the variables according to the empirical variance. For each
$p$ we delete $10 \%$ of the data, run MissPALasso and MissGLasso
ten times on a decreasing grid (on the log-scale) of $\lambda$ values with
thirty grid points. For a fixed $\lambda$ we stop the algorithm if the relative
change in imputation satisfies,
\begin{eqnarray*}
\frac{\|\hat{\X}^{(r+1)}-\hat{\X}^{(r)}\|^2}{\|\hat{\X}^{(r+1)}\|^2}\leq 10^{-5}.
\end{eqnarray*} 
In Figure~\ref{fig:timing} the CPU times in seconds are plotted for various
values of $p$ in the two settings. As shown, with MissPALasso we are
typically able to solve a problem of size $p=100$ in about 9 seconds and a
problem of size $p=500$ in about 400 seconds. For MissGLasso these times are highly increased to 27 and 4300 seconds
respectively. Furthermore, we can see that MissPALasso has much smaller
variability in runtimes. The graphical Lasso algorithm
\citep{friedman2007sic} and therefore MissGLasso have computational
complexity $\mathcal{O}(p^3)$, whereas the complexity of MissPALasso is
considerably smaller. We note that the matrix inversion
in the M-Step of MissPA is replaced by soft-thresholding in MissPALasso. In sparse
settings soft-thresholding as well as matrix
multiplications can be evaluated very quickly as the regression
coefficients $B_{m_k|o_k}$ contain many zeros.
The exact complexity of MissPALasso depends
on the fraction of missing values and also on the sparsity of the problem.

\section{Theory}\label{sec:theory}
A key characteristic of our MissPA (Algorithm 1 in
Section~\ref{sec:patternwise}) is that the E-Step is only performed on those samples
belonging to a single pattern. We already mentioned the close
connection to the incremental EM introduced by \cite{incremental}. In
fact, if the density of $\X^{k}$, $k \in \{1,\ldots,s\}$, is denoted by
$\PP_{\Sigma}(\X^{k})=\prod_{i\in \mathcal{I}_k}p(x_{i};\Sigma)$
then the negative variational free energy \citep{incremental,jordan1999} equals
\begin{eqnarray}\label{eq:var.energy}
\F[\Sigma\|\Psi_{1},\ldots,\Psi_{s}]&=&\sum_{k=1}^{s}
\big(\E_{\Psi_{k}}[\log \PP_{\Sigma}(\X^k)|\X^{k}_{o_k}]+\HH_k[\Psi_{k}]\big).
\end{eqnarray}
Here, $\Psi_{k}=\left(B_{k,m_k|o_k},\Sigma_{k,m_k|o_k}\right)$ denotes
the regression parameter of the latent distribution
\begin{eqnarray*}
\PP_{\Psi_k}(\X^{k}_{m_k}|\X^{k}_{o_k})&=&\prod_{i\in \mathcal{I}_k}p(x_{i,m_k}|x_{i,o_k};B_{k,m_k|o_k},\Sigma_{k,m_k|o_k})
\end{eqnarray*}
and $\HH_k[\Psi_k]=-\E_{\Psi_k}[\log
\PP_{\Psi_k}(\X_{m_k}^k|\X_{o_k}^k)|\X^{k}_{o_k}]$ is the
entropy. An iterative procedure alternating between
maximization of $\F$ with respect to $\Sigma$
\begin{eqnarray*}
 \hat{\Sigma}&=&\argmax_{\Sigma} \F[\Sigma\|\Psi_{1},\ldots,\Psi_{s}]\\&=&\frac{1}{n}\sum_{k=1}^s\E_{\Psi_{k}}[^t\X^k\X^k|\X^k_{o_k}]=:\frac{1}{n}\calT,
\end{eqnarray*}
and maximizing $\F$ with respect to $\Psi_{k}$
\begin{eqnarray*}
 (\hat B_{k,m_k|o_k},\hat \Sigma_{k,m_k|o_k})&=&\argmax_{\Psi_{k}}\F[\hat
 \Sigma\|\Psi_{1},\ldots,\Psi_{s}]\\
&=&\argmax_{\Psi_{k}}\E_{\Psi_{{k}}}[\log
\PP_{\hat\Sigma}(\X^k)|\X^{k}_{o_k}]+\HH_k[\Psi_{k}]\\
&=&\left(\calT_{m_k,o_k}\calT_{o_k,o_k}^{-1},\frac{1}{n}(\calT_{m_k,m_k}-\calT_{m_k,o_k}\calT_{o_k,o_k}^{-1}\calT_{o_k,m_k})\right)
  \end{eqnarray*}
is equivalent to our Algorithm~1 with $\calT^{-k}$ replaced by $\calT$
(see M-Step2 in Section~\ref{sec:patternwise}).  Alternating maximization of (\ref{eq:var.energy}) is a GAM
procedure in the sense of \cite{gam2005}. Within their framework convergence to a
stationary point of the observed log-likelihood can be
shown. 

Unfortunately, the MissPA algorithm does not quite fit into the
GAM formulation. As already mentioned MissPA extends the
standard EM
also in another way namely by using for each pattern a different
complete data space (for each pattern $k$ only those samples are augmented which do not belong to
pattern $k$). MissPA is related to the SAGE algorithm
\citep{sage1994} as follows: Consider the parameter of interest $\Sigma$
in the parameterization
$\theta=\left(\Sigma_{o_k},B_{m_k|o_k},\Sigma_{m_k|o_k}\right)$
introduced in Section~\ref{sec:setup}, equation~(\ref{eq:transf}). From
$$\PP_{\theta}(\X_{\rm obs},\X^{-k})=\PP_{\theta}(\X_{\rm obs}|\X^{-k})\PP_{\theta}(\X^{-k})$$
and from observing that $\PP_{\theta}(\X_{\rm
  obs}|\X^{-k})=\PP_{\Sigma_{o_k}}(\X_{o_k})$ we conclude that $\X^{-k}$ is an
admissible hidden-data space with respect to
$(B_{m_k|o_k},\Sigma_{m_k|o_k})$ in the sense of
\cite{sage1994}. The M-Step of MissPA then maximizes a conditional
expectation of the log-likelihood $\log\PP_{\theta}(\X^{-k})$ with
respect to  $(B_{m_k|o_k},\Sigma_{m_k|o_k})$. Different from a SAGE
algorithm is the conditional distribution involved in the
expectation. Our algorithm updates after each M-Step only the
conditional distribution for a single pattern. As a consequence, we do not need to compute estimates for
$\Sigma_{o_k}$. 


In summary, the MissPA algorithm has similarities with a GAM and a
SAGE procedure. However, neither the SAGE nor the GAM framework fit our proposed
MissPA. In the next
section we provide theory which
justifies alternating between complete data spaces \emph{and} incrementally performing
the E-Step. In particular, we prove convergence to a stationary point
of the observed log-likelihood.

\subsection{Convergence Theory for the MissPA Algorithm}\label{sec:theory.misspa}
\paragraph{Pattern-Depending Lower Bounds}
Denote the density of $\X^{k}$, $k \in \{1,\ldots,s\}$, by
$\PP_{\Sigma}(\X^{k})=\prod_{i\in \mathcal{I}_k}p(x_{i};\Sigma)$ and define for $k,l \in \{1,\ldots,s\}$
\begin{eqnarray*}
\PP_{\Sigma}(\X^{l}_{o_k})&=&\prod_{i\in \mathcal{I}_l}p(x_{i,o_k};\Sigma_{o_k})\quad \textrm{and}\\
\PP_{\Sigma}(\X^{l}_{m_k}|\X^{l}_{o_k})&=&\prod_{i\in \mathcal{I}_l}p(x_{i,m_k}|x_{i,o_k};B_{m_k|o_k},\Sigma_{m_k|o_k}).
\end{eqnarray*}


Set $\{\Sigma_l\}_{l\neq
  k}=(\Sigma_1,\ldots,\Sigma_{k-1},\Sigma_{k+1},\ldots,\Sigma_{s})$ and consider for $k=1,\ldots,s$ 
\begin{eqnarray}
\F_{k}[\Sigma_k||\{\Sigma_l\}_{l\neq
  k}]&=&\log \PP_{\Sigma_k}(\X^k_{o_k})+\sum_{l\neq k}
\big(\E_{\Sigma_{l}}[\log \PP_{\Sigma_k}(\X^l)|\X^{l}_{o_l}]+\HH_l[\Sigma_{l}]\big).\nonumber
\end{eqnarray}
Here $\HH_l[\tilde{\Sigma}]=-\E_{\tilde{\Sigma}}[\log
\PP_{\tilde{\Sigma}}(\X_{m_l}^l|\X_{o_l}^l)|\X^{l}_{o_l}]$ denotes the entropy. Note
that $\F_k$ is defined for fixed observed data $\X_{\mathrm{obs}}$. The
subscript $k$  highlights the dependence on
the pattern $k$. Furthermore, for fixed $\X_{\mathrm{obs}}$ and fixed $k$, $\F_k$ is a function in the parameters $(\Sigma_1,\ldots,\Sigma_s)$.
As a further tool we write the Kullback-Leibler divergence in the
following form:
\begin{eqnarray}\label{eq:kldivergence}
\D_l[\tilde{\Sigma}||\Sigma]=\E_{\tilde{\Sigma}}[-\log\big(\PP_{\Sigma}(\X^l_{m_l}|\X^l_{o_l})/\PP_{\tilde{\Sigma}}(\X^l_{m_l}|\X^l_{o_l})\big)|\X^{l}_{o_l}].
\end{eqnarray}
An important property of the Kullback-Leibler
divergence is its non-negativity:
\begin{eqnarray}
&&\D_l[\tilde{\Sigma}||\Sigma]\geq 0, \quad\textrm{with equality if and
  only if} \nonumber\\
&&\PP_{\tilde{\Sigma}}(\X^l_{m_l}|\X^l_{o_l})=\PP_{\Sigma}(\X^l_{m_l}|\X^l_{o_l}).\nonumber
\end{eqnarray}

A simple calculation shows that
\begin{eqnarray}
&&\E_{\tilde{\Sigma}}[\log \PP_{\Sigma}(\X^l)|\X^{l}_{o_l}]+\HH_l[\tilde{\Sigma}]=-\D_l[\tilde{\Sigma}||\Sigma]+\log
\PP_{\Sigma}(\X^l_{o_l})\label{eq:inform2}
\end{eqnarray}
and $\F_{k}[\Sigma_k||\{\Sigma_l\}_{l\neq k}]$ can be written as
\begin{eqnarray}\label{eq:crit2}
\F_{k}[\Sigma_k||\{\Sigma_l\}_{l\neq k}]&=&\ell(\Sigma_k;\X_{\mathrm{obs}})-\sum_{l\neq k}\D_l[\Sigma_l||\Sigma_k].
\end{eqnarray}
In particular, for fixed values of $\{\Sigma_l\}_{l\neq k}$, $\F_{k}[\,\cdot\,||\{\Sigma_l\}_{l\neq k}]$ lower bounds the observed
log-likelihood $\ell(\,\cdot\, ;\X_{\mathrm{obs}})$ due to the non-negativity of the
Kullback-Leibler divergence.

\paragraph{Optimization Transfer to Pattern-Depending
  Lower Bounds}
We give now an alternative description of the MissPA algorithm.
In cycle $r+1$ through all patterns, generate
$(\Sigma_1^{r+1},\ldots,\Sigma_s^{r+1})$ given $(\Sigma_1^{r},\ldots,\Sigma_s^{r})$ according to
\begin{eqnarray}\label{eq:alg1abstract}
&&\Sigma_k^{r+1}=\argmax_{\Sigma}\F_k[\Sigma||\ZZ_k^{r+1}],\quad
k=1,\ldots,s,
\end{eqnarray}
with $\ZZ_k^{r+1}=(\Sigma_1^{r+1},\ldots,\Sigma_{k-1}^{r+1},\Sigma_{k+1}^r,\ldots,\Sigma_s^r).$

We have
\begin{eqnarray*}
\F_k[\Sigma||\ZZ_k^{r+1}]&=&\log \PP_{\Sigma}(\X^k_{o_k})+\sum_{l<
  k}\left(\E_{\Sigma_l^{r+1}}[\log
\PP_{\Sigma}(\X^{l})|\X^{l}_{o_l}]+\HH_l[\Sigma_l^{r+1}]\right)\\
&&+\sum_{l>k}\left(\E_{\Sigma_l^{r}}[\log \PP_{\Sigma}(\X^{l})|\X^{l}_{o_l}]+\HH_l[\Sigma_l^{r}]\right).\\ 
\end{eqnarray*}
The entropy terms do not depend on the optimization parameter $\Sigma$,
therefore,
\begin{align*}
\F_k[\Sigma||\ZZ_k^{r+1}]&=\textrm{const}+\log \PP_{\Sigma}(\X^k_{o_k})+\sum_{l<
  k}\E_{\Sigma_l^{r+1}}[\log
\PP_{\Sigma}(\X^{l})|\X^{l}_{o_l}]+\sum_{l>k}\E_{\Sigma_l^{r}}[\log
\PP_{\Sigma}(\X^{l})|\X^{l}_{o_l}].
\end{align*}
Using the factorization $\log \PP_{\Sigma}(\X^{l})=\log
\PP(\X_{o_{k}}^{l};\Sigma_{o_k})+\log \PP(\X_{m_k}^l|\X_{o_k}^{l};B_{m_k|o_k},\Sigma_{m_k|o_k})$ (for all $l\neq k$),
and separate maximization with respect to $\Sigma_{o_k}$ and
$(B_{m_k|o_k},\Sigma_{m_k|o_k})$ we end up with the expressions from the
M-Step of MissPA. Summarizing the above, we have recovered the M-Step
as a maximization of $\F_k[\Sigma||\ZZ_k^{r+1}]$ which is a lower bound of
the observed log-likelihood. Or in the language of \cite{lange2000},
optimization of $\ell(\,\cdot\,;\X_{\mathrm{obs}})$ is transferred to the surrogate objective $\F_k[\,\cdot\,||\ZZ_k^{r+1}]$. 

There is still an important piece missing: In M-Step $k$ of cycle $r+1$ we
are maximizing $\F_k[\;\cdot\;||\ZZ_k^{r+1}]$ whereas in the following
M-Step ($k+1$) we optimize $\F_{k+1}[\;\cdot\;||\ZZ_{k+1}^{r+1}]$. In order
for the algorithm to make progress, it is essential that
$\F_{k+1}[\;\cdot\;||\ZZ_{k+1}^{r+1}]$ attains higher values than its
predecessor $\F_{k}[\;\cdot\;||\ZZ_{k}^{r+1}]$. In this sense the following
proposition is crucial.
\\
\begin{prop}\label{eq:prop1}
For $r=0,1,2,\ldots$ we have that
\begin{eqnarray*}
&&\F_s[\Sigma_s^{r}||\ZZ_s^{r}]\leq\F_1[\Sigma_s^{r}||\ZZ_1^{r+1}],\quad \textrm{and}\\
&&\F_{k}[\Sigma_k^{r+1}||\ZZ_k^{r+1}]\leq
\F_{k+1}[\Sigma_k^{r+1}||\ZZ_{k+1}^{r+1}]\quad\textrm{for
  $k=1,\ldots,s-1$}.
\end{eqnarray*}
\end{prop}

{\bf Proof}.
{
We have,
$$\F_{k}[\Sigma_k^{r+1}||\ZZ_k^{r+1}]=\log
\PP_{\Sigma_k^{r+1}}(\X^k_{o_k})+\E_{\Sigma_{k+1}^{r}}[\log
\PP_{\Sigma_k^{r+1}}(\X^{k+1})|\X_{o_{k+1}}^{k+1}]+\HH_{k+1}[\Sigma_{k+1}^r]+\textrm{rest}$$
and
$$\F_{k+1}[\Sigma_k^{r+1}||\ZZ_{k+1}^{r+1}]=\log
\PP_{\Sigma_k^{r+1}}(\X^{k+1}_{o_{k+1}})+\E_{\Sigma_{k}^{r+1}}[\log
\PP_{\Sigma_k^{r+1}}(\X^{k})|\X_{o_{k}}^{k}]+\HH_k[\Sigma_{k}^{r+1}]+\textrm{rest}$$
where
$$\textrm{rest}=\sum_{l< k}\big(\E_{\Sigma_l^{r+1}}[\log \PP_{\Sigma_k^{r+1}}(\X^l)|\X^{l}_{o_l}]+\HH_l[\Sigma_l^{r+1}]\big)+\sum_{l> k+1}\big(\E_{\Sigma_l^{r}}[\log \PP_{\Sigma_k^{r+1}}(\X^l)|\X^{l}_{o_l}]+\HH_l[\Sigma_l^{r}]\big).$$
Furthermore, using (\ref{eq:inform2}) and noting that
$\D_k[\Sigma_{k}^{r+1}||\Sigma_{k}^{r+1}]=0$, we obtain

\begin{align*}
\F_{k}[\Sigma_k^{r+1}||\ZZ_k^{r+1}]-\F_{k+1}[\Sigma_k^{r+1}||\ZZ_{k+1}^{r+1}]&=\D_k[\Sigma_{k}^{r+1}||\Sigma_{k}^{r+1}]-\D_{k+1}[\Sigma_{k+1}^r||\Sigma_{k}^{r+1}]\\
&=
-\D_{k+1}[\Sigma_{k+1}^r||\Sigma_{k}^{r+1}]\leq
0.
\end{align*}
Note that equality holds if and only if
$\PP_{\Sigma_{k}^{r+1}}(\X^{k+1}_{m_{k+1}}|\X^{k+1}_{o_{k+1}})=\PP_{\Sigma_{k+1}^{r}}(\X^{k+1}_{m_{k+1}}|\X^{k+1}_{o_{k+1}})$.}\hfill$\Box$

In light of Proposition~\ref{eq:prop1} it is clear that
(\ref{eq:alg1abstract}) generates a monotonely increasing sequence of the
form:
\begin{align*}
&\F_s[\Sigma_s^0||\ZZ_s^0]\leq\F_1[\Sigma_s^0||\ZZ_1^1]\leq\F_1[\Sigma_1^1||\ZZ_1^1]\leq
\F_2[\Sigma_1^{1}||\ZZ_2^{1}]\leq\F_{2}[\Sigma_2^{1}||\ZZ_{2}^{1}]\leq
\cdots\\
&\cdots\leq\F_k[\Sigma_k^{r+1}||\ZZ_k^{r+1}]\leq\F_{k+1}[\Sigma_k^{r+1}||\ZZ_{k+1}^{r+1}]\leq\F_{k+1}[\Sigma_{k+1}^{r+1}||\ZZ_{k+1}^{r+1}]\leq\cdots
\end{align*}
For example, we can deduce that
$\{\F_{s}[\Sigma_s^{r}||\ZZ_{s}^{r}]\}_{r=0,1,2,\ldots}$ is a monotone
  increasing sequence in $r$.
\paragraph{Convergence to Stationary Points}
Ideally we would like to show that a limit point of the sequence generated by the MissPA
algorithm is a global maximum of $\ell(\Sigma;\X_{\mathrm{obs}})$. Unfortunately, this is too ambitious because for general missing data patterns the observed log-likelihood is a non-concave
function with several local maxima. Thus, the most we can expect
is that our algorithm converges to a stationary point. This is ensured by
the following theorem which is proved in the Appendix.
\\
\begin{theorem}\label{thm:convergence}
Assume that 
$\mathcal{K}=\{(\Sigma_1,\ldots,\Sigma_s):\F_{s}[\Sigma_s||\Sigma_1,\ldots,\Sigma_{s-1}]\geq\F_{s}[\Sigma_s^{0}||\ZZ_s^0]\}$
is compact. Then every limit point $\bar{\Sigma}_s$ of $\{\Sigma_s^{r}\}_{r=0,1,2,\ldots}$ is a stationary point of $\ell(\;\cdot\;;\X_{\mathrm{obs}})$.
\end{theorem}

\section{Discussion and Extensions}
We have presented the novel Missingness Pattern Alternating
maximization algorithm (MissPA) for maximizing the observed
log-likelihood for a multivariate normal data matrix with missing
values. Simplified, our algorithm iteratively cycles through the different missingness
patterns, performs multivariate regressions of the missing on the
observed variables and uses these regression coefficients for partial
imputation of the missing values. We argued theoretically and gave numerical examples
showing that our procedure is computationally more efficient than the standard EM
algorithm. Furthermore, we analyze the numerical properties using
non-standard arguments and prove that solutions of our algorithm converge to stationary points
of the observed log-likelihood.

In a high-dimensional setup with $p\gg n$ the regression interpretation opens up the door to do regularization by replacing least
squares regressions with Lasso analogues. Our proposed algorithm, the MissPALasso, performs
a coordinate descent approximation of the corresponding Lasso problem in
order to gain speed. On simulated and four real datasets (all from
genomics) we demonstrate that MissPALasso outperforms other imputation techniques such as k-nearest neighbors imputation, nuclear norm
minimization or a penalized likelihood approach with an $\ell_1$-penalty on
the inverse covariance matrix.

Even though MissPALasso performs well on simulated and real data
it is a ``heuristic'' motivated by the previously developed
MissPA and by the wish to have sparse regression coefficients
for imputation. It is unclear which objective function is
optimized by MissPALasso. The comments of two referees on this
point made us
thinking of another way of imposing sparsity in the regression
coefficients:
Consider a penalized variational free energy  
\begin{eqnarray}\label{eq:pen.var.energy}
&-\F[\Sigma\|\Psi_{1},\ldots,\Psi_{s}]+\sum_{k=1}^s\mathrm{Pen}_{\lambda}(\Psi_k),
\end{eqnarray} 
with $\F[\Sigma\|\Psi_{1},\ldots,\Psi_{s}]$ defined in
equation~(\ref{eq:var.energy}) and
$\mathrm{Pen}_{\lambda}(\Psi_k)=\lambda\|B_{k,m_k|o_k}\|_1$. A small
calculation shows that alternating minimization of (\ref{eq:pen.var.energy}) with respect to
$\Sigma$ and $\Psi_k$ leads to an algorithm which is similar but
different from MissPALasso.
In fact, minimizing (\ref{eq:pen.var.energy}) with respect to
$\Sigma_{k,m_k|o_k}$ and $B_{k,m_k|o_k}$ gives
$\hat\Sigma_{k,m_k|o_k}=\Sigma_{m_k|o_k}$
and $\hat B_{k,m_k|o_k}$ satisfies the subgradient equation
$$0=\left(\Omega_{m_k,m_k} \hat B_{k,m_k|o_k} -\Omega_{m_k,o_k}\right)^t\!\X_{o_k}^k
\X_{o_k}^k+\lambda \Gamma(\hat B_{k,m_k|o_k}),$$
where $\Gamma(x)$ is the subgradient of $|x|$, applied componentwise
to the elements of a matrix and $\Omega=\Sigma^{-1}$. The formulation (\ref{eq:pen.var.energy})
looks very compelling and can motivate new algorithms for missing
data imputation. For example in the context of the Netflix problem where
the fraction of the missing values and the number of customers is huge one might impose constraints on the parameters
$\Sigma_{k,m_k|o_k}$ and employ a stochastic gradient descent
optimization strategy to solve (\ref{eq:pen.var.energy}).

\appendix
\section{Proof of Theorem \ref{thm:convergence}}
{\bf Proof.}
First, note that the sequence $\{(\Sigma_1^{r},\ldots,\Sigma_s^{r})\}_{r=0,1,2,\ldots}$ lies in the compact
  set $\mathcal{K}$. Now, let $\Sigma_s^{r_j}$ be a subsequence converging to $\bar{\Sigma}_s$
  as $j\rightarrow \infty$. By invoking compactness, we can assume w.l.o.g (by restricting to a subsequence) that $(\Sigma_1^{r_j},\ldots,\Sigma_s^{r_j})\rightarrow (\bar{\Sigma}_1,\ldots,\bar{\Sigma}_s)$. 

As a direct consequence of the monotonicity of the sequence $\{\F_{s}[\Sigma_s^{r}||\ZZ_{s}^{r}]\}_{r=0,1,2,\ldots}$ we obtain 
$$\lim_r\F_{s}[\Sigma_s^{r}||\ZZ_{s}^{r}]=\F_s[\bar{\Sigma}_s||\bar{\Sigma}_1,\ldots,\bar{\Sigma}_{s-1}]\equiv\bar{\F}.$$
From (\ref{eq:alg1abstract}) and Proposition~\ref{eq:prop1}, for $k=1,\ldots,s-1$ and $r=0,1,2,\ldots,$ the following ``sandwich''-formulae hold:
\begin{eqnarray*}
&&\F_{s}[\Sigma_s^{r}||\ZZ_{s}^{r}]\leq\F_{1}[\Sigma_s^{r}||\ZZ_{1}^{r+1}]\leq\F_{1}[\Sigma_{1}^{r+1}||\ZZ_{1}^{r+1}]\leq\F_{s}[\Sigma_s^{r+1}||\ZZ_{s}^{r+1}],\\
&&\F_{s}[\Sigma_s^{r}||\ZZ_{s}^{r}]\leq\F_{k+1}[\Sigma_k^{r+1}||\ZZ_{k+1}^{r+1}]\leq\F_{k+1}[\Sigma_{k+1}^{r+1}||\ZZ_{k+1}^{r+1}]\leq\F_{s}[\Sigma_s^{r+1}||\ZZ_{s}^{r+1}].
\end{eqnarray*}
As a consequence we have for $k=1,\ldots,s-1$
\begin{eqnarray}
&&\lim_r\F_{1}[\Sigma_s^{r}||\ZZ_{1}^{r+1}]=\lim_r\F_{1}[\Sigma_1^{r+1}||\ZZ_{1}^{r+1}]=\bar{\F}\quad\textrm{and}\label{eq:sandwich1}\\
&&\lim_r\F_{k+1}[\Sigma_k^{r+1}||\ZZ_{k+1}^{r+1}]=\lim_r\F_{k+1}[\Sigma_{k+1}^{r+1}||\ZZ_{k+1}^{r+1}]=\bar{\F}.\label{eq:sandwich2}
\end{eqnarray}
Now consider the sequence $(\Sigma_1^{r_j+1},\ldots,\Sigma_s^{r_j+1})$. By compactness of $\mathcal{K}$ this sequence converges w.l.o.g to some $(\Sigma^*_1,\ldots,\Sigma^*_s)$. We now show by induction that 
\[\bar{\Sigma}_s=\Sigma^*_1=\ldots=\Sigma^*_s.\]
From the $1$st M-Step of cycle $r_j+1$ we have
\[\F_{1}[\Sigma_1^{r_j+1}||\ZZ_1^{r_j+1}]\geq\F_{1}[\Sigma||\ZZ_1^{r_j+1}]\quad \textrm{for all}\quad \Sigma.\]
Taking the limit $j\rightarrow\infty$ we get:
\[\F_{1}[\Sigma^{*}_1||\{\bar{\Sigma}_l\}_{l>1}]\geq
\F_{1}[\Sigma|\{\bar{\Sigma}_l\}_{l>1}] \quad \textrm{for all}\quad \Sigma.\]
In particular,  $\Sigma^{*}_1$ is the (unique) maximizer of $\F_{1}[\,\cdot\,||\{\bar{\Sigma}_l\}_{l>1}]$. Assuming $\Sigma^*_1\neq\bar{\Sigma}_s$ would imply
\[\F_{1}[\Sigma^*_1||\{\bar{\Sigma}_l\}_{l>1}]>
\F_{1}[\bar{\Sigma}_s||\{\bar{\Sigma}_l\}_{l>1}].\]
But this contradicts $\F_{1}[\Sigma^*_1||\{\bar{\Sigma}_l\}_{l>1}]=
\F_{1}[\bar{\Sigma}_s||\{\bar{\Sigma}_l\}_{l>1}]=\bar{\F}$, which holds by (\ref{eq:sandwich1}). Therefore we obtain $\Sigma_1^*=\bar{\Sigma}_s$. 

Assume that we have proven $\Sigma^*_1=\ldots=\Sigma^*_k=\bar{\Sigma}_s$. We will show that $\Sigma_{k+1}^*=\bar{\Sigma}_s$. From the k+1st M-Step in cycle $r_j+1$:
\[\F_{k+1}[\Sigma_{k+1}^{r_j+1}||\ZZ_{k+1}^{r_j+1}]\geq \F_{k+1}[\Sigma||\ZZ_{k+1}^{r_j+1}] \quad\textrm{for all}\quad \Sigma.\]
Taking the limit for $j\rightarrow \infty$, we conclude that $\Sigma^*_{k+1}$ is
the (unique) maximizer of $$\F_{k+1}[\,\cdot\,||\{\Sigma^*_l\}_{l<
  k+1},\{\bar{\Sigma}_l\}_{l>k+1}].$$
From (\ref{eq:sandwich2}),
\[\F_{k+1}[\Sigma^*_{k+1}||\{\Sigma^{*}_l\}_{l<
  k+1},\{\bar{\Sigma}_l\}_{l>k+1}]=\F_{k+1}[\Sigma^{*}_k||\{\Sigma^{*}_l\}_{l<
  k+1},\{\bar{\Sigma}_l\}_{l>k+1}]=\bar{\F},\]
and therefore $\Sigma^*_{k+1}$ must be equal to $\Sigma^{*}_k$. By
induction we have $\Sigma^*_{k}= \bar{\Sigma}_s$ and so we proved that
$\Sigma^*_{k+1}=\bar{\Sigma}_s$ holds.

Finally, we show stationarity of $\bar{\Sigma}_s$. Invoking (\ref{eq:crit2}) we can write
\[
\F_{s}[\Sigma||\bar{\Sigma}_s,\ldots,\bar{\Sigma}_s]=\ell(\Sigma;\X_{\mathrm{obs}})-\sum_{l=1}^{s-1}\D_l[\bar{\Sigma}_s||\Sigma].\]
Note that
\[\deriv
\D_l[\bar{\Sigma}_s||\Sigma]\bigg|_{\bar{\Sigma}_s}=0.
\]
Furthermore, as $\Sigma_s^{r_j+1}$ maximizes
$\F_{s}[\Sigma||\Sigma_1^{r_j+1},\ldots,\Sigma_{s-1}^{r_j+1}]$, we get in
the limit as $j\rightarrow \infty$
\[\deriv\F_{s}[\Sigma|\bar{\Sigma}_s,\ldots,\bar{\Sigma}_s]\bigg|_{\bar{\Sigma}_s}=\deriv\F_{s}[\Sigma||\Sigma^*_1,\ldots,\Sigma^*_{s-1}]\bigg|_{\Sigma^*_s}=0.\] 
Therefore, we conclude that $\deriv\ell(\Sigma;\X_{\mathrm{obs}})\Big|_{\bar{\Sigma}_s}=0$.
\hfill$\Box$    
\vskip 0.2in


\begin{thebibliography}{30}
\expandafter\ifx\csname natexlab\endcsname\relax\def\natexlab#1{#1}\fi
\expandafter\ifx\csname url\endcsname\relax
  \def\url#1{\texttt{#1}}\fi
\expandafter\ifx\csname urlprefix\endcsname\relax\def\urlprefix{URL }\fi

\bibitem[{Aittokallio(2010)}]{aittokallio2009}
Aittokallio, T. (2010) Dealing with missing values in large-scale studies:
  microarray data imputation and beyond.
\newblock \emph{Briefings in Bioinformatics}, \textbf{11}, 253--264.

\bibitem[{Alizadeh \emph{et~al.}(2000)Alizadeh, Eisen, Davis, Ma, Lossos,
  Rosenwald, Boldrick, Sabet, Tran, Yu, Powell, Yang, Marti, Moore, Hudson, Lu,
  Lewis, Tibshirani, Sherlock, Chan, Greiner, Weisenburger, Armitage, Warnke,
  Levy, Wilson, Grever, Byrd, Botstein, Brown and Staudt}]{lymphoma2000}
Alizadeh, A., Eisen, M., Davis, R., Ma, C., Lossos, I., Rosenwald, A.,
  Boldrick, J., Sabet, H., Tran, T., Yu, X., Powell, J., Yang, L., Marti, G.,
  Moore, T., Hudson, J., Lu, L., Lewis, D., Tibshirani, R., Sherlock, G., Chan,
  W., Greiner, T., Weisenburger, D., Armitage, J., Warnke, R., Levy, R.,
  Wilson, W., Grever, M.~R., Byrd, J., Botstein, D., Brown, P. and Staudt, L.
  (2000) Distinct types of diffuse large b-cell lymphoma identified by gene
  expression profiling.
\newblock \emph{Nature}, \textbf{403}, 503--511.

\bibitem[{Allen and Tibshirani(2010)}]{transposable2010}
Allen, G. and Tibshirani, R. (2010) Transposable regularized covariance models
  with an application to missing data imputation.
\newblock \emph{Annals of Applied Statistics}, \textbf{4}, 764--790.

\bibitem[{Alon \emph{et~al.}(1999)Alon, Barkai, Notterman, Gishdagger,
  Ybarradagger, Mackdagger and Levine}]{alon99}
Alon, U., Barkai, N., Notterman, D., Gishdagger, K., Ybarradagger, S.,
  Mackdagger, D. and Levine, A. (1999) Broad patterns of gene expression
  revealed by clustering analysis of tumor and normal colon tissues probed by
  oligonucleotide arrays.
\newblock \emph{Proceedings of the National Academy of Sciences of the United
  States of America}, \textbf{96}, 6745--6750.

\bibitem[{Cai \emph{et~al.}(2010)Cai, Cand\`{e}s and Shen}]{svt2008}
Cai, J.-F., Cand\`{e}s, E. and Shen, Z. (2010) A singular value thresholding
  algorithm for matrix completion.
\newblock \emph{SIAM Journal on Optimization}, \textbf{20}, 1956--1982.

\bibitem[{Cand{\`e}s and Recht(2009)}]{candesandrecht2009}
Cand{\`e}s, E. and Recht, B. (2009) Exact matrix completion via convex
  optimization.
\newblock \emph{Foundations of Computational Mathematics}, \textbf{9},
  717--772.

\bibitem[{Cand{\`e}s and Tao(2010)}]{candesandtao2009}
Cand{\`e}s, E. and Tao, T. (2010) The power of convex relaxation: Near-optimal
  matrix completion.
\newblock \emph{IEEE Transactions on Information Theory}, \textbf{56}.

\bibitem[{Dempster \emph{et~al.}(1977)Dempster, Laird and Rubin}]{Dempster}
Dempster, A., Laird, N. and Rubin, D. (1977) Maximum likelihood from incomplete
  data via the {EM} algorithm.
\newblock \emph{Journal of the Royal Statistical Society, Series B},
  \textbf{39}, 1--38.

\bibitem[{Fessler and Hero(1994)}]{sage1994}
Fessler, J. and Hero, A. (1994) Space-alternating generalized
  {E}xpectation-{M}aximization algorithm.
\newblock \emph{IEEE Transactions on Signal Processing}, \textbf{42},
  2664--2677.

\bibitem[{Friedman \emph{et~al.}(2008)Friedman, Hastie and
  Tibshirani}]{friedman2007sic}
Friedman, J., Hastie, T. and Tibshirani, R. (2008) Sparse inverse covariance
  estimation with the graphical {L}asso.
\newblock \emph{Biostatistics}, \textbf{9}, 432--441.

\bibitem[{Friedman \emph{et~al.}(2010)Friedman, Hastie and
  Tibshirani}]{friedmanetal08}
Friedman, J., Hastie, T. and Tibshirani, R. (2010) Regularization paths for
  generalized linear models via coordinate descent.
\newblock \emph{Journal of Statistical Software}, \textbf{33}, 1--22.

\bibitem[{Gunawardana and Byrne(2005)}]{gam2005}
Gunawardana, A. and Byrne, W. (2005) Convergence theorems for generalized
  alternating minimization procedures.
\newblock \emph{Journal of Machine Learning Research}, \textbf{6}, 2049--2073.

\bibitem[{Jordan \emph{et~al.}(1999)Jordan, Ghahramani, Jaakkola and
  Saul}]{jordan1999}
Jordan, M., Ghahramani, Z., Jaakkola, T. and Saul, L. (1999) An introduction to
  variational methods for graphical models.
\newblock \emph{Machine Learning}, \textbf{37}, 183--233.

\bibitem[{Keshavan \emph{et~al.}(2010)Keshavan, Oh and
  Montanari}]{keshavan2009}
Keshavan, R., Oh, S. and Montanari, A. (2010) Matrix completion from a few
  entries.
\newblock \emph{IEEE Transactions on Information Theory}, \textbf{56}.

\bibitem[{Kim \emph{et~al.}(2006)Kim, Golub and Park}]{localleastsquares2006}
Kim, H., Golub, G. and Park, H. (2006) Missing value estimation for {DNA}
  microarray gene expression data: local least squares imputation.
\newblock \emph{Bioinformatics}, \textbf{22}, 1410--1411.

\bibitem[{Lange \emph{et~al.}(2000)Lange, Hunter and Yang}]{lange2000}
Lange, K., Hunter, D. and Yang, I. (2000) Optimization transfer using surrogate
  objective functions.
\newblock \emph{Journal of Computational and Graphical Statistics}, \textbf{9},
  1--20.

\bibitem[{Little and Rubin(1987)}]{LittleRubin}
Little, R. and Rubin, D. (1987) \emph{Statistical Analysis with Missing Data}.
\newblock Series in Probability and Mathematical Statistics, Wiley.

\bibitem[{Mazumder \emph{et~al.}(2010)Mazumder, Hastie and
  Tibshirani}]{softimpute2008}
Mazumder, R., Hastie, T. and Tibshirani, R. (2010) Spectral regularization
  algorithms for learning large incomplete matrices.
\newblock \emph{Journal of Machine Learning Research}, \textbf{99}, 2287--2322.

\bibitem[{Neal and Hinton(1998)}]{incremental}
Neal, R. and Hinton, G. (1998) A view of the {EM} algorithm that justifies
  incremental, sparse, and other variants.
\newblock In \emph{Learning in Graphical Models},  355--368. Kluwer Academic
  Publishers.

\bibitem[{Ng and McLachlan(2003)}]{McLachlanNg2003}
Ng, S. and McLachlan, G. (2003) On the choice of the number of blocks with the
  incremental {EM} algorithm for the fitting of normal mixtures.
\newblock \emph{Statistics and Computing}, \textbf{13}, 45--55.

\bibitem[{Nowlan(1991)}]{nowlan91}
Nowlan, S. (1991) \emph{Soft Competitive Adaptation: Neural Network Learning
  Algorithms based on Fitting Statistical Mixtures}.
\newblock Ph.D. thesis, School of Computer Science, Carnegie Mellon University,
  Pittsburgh.

\bibitem[{Oba \emph{et~al.}(2003)Oba, Sato, Takemasa, Monden, Matsubara and
  Ishii}]{bpcaimpute}
Oba, S., Sato, M.-A., Takemasa, I., Monden, M., Matsubara, K.-I. and Ishii, S.
  (2003) A {B}ayesian missing value estimation method for gene expression
  profile data.
\newblock \emph{Bioinformatics}, \textbf{19}, 2088--2096.

\bibitem[{Schafer(1997)}]{Schafer}
Schafer, J. (1997) \emph{Analysis of Incomplete Multivariate Data}.
\newblock Monographs on Statistics and Applied Probability 72, Chapman and
  Hall.

\bibitem[{Spellman \emph{et~al.}(1998)Spellman, Sherlock, Zhang, Iyer, Anders,
  Eisen, Brown, Botstein and Futcher}]{spellman98}
Spellman, P., Sherlock, G., Zhang, M., Iyer, V., Anders, K., Eisen, M., Brown,
  P., Botstein, D. and Futcher, B. (1998) Comprehensive identification of cell
  cycle-regulated genes of the yeast {S}accharomyces cerevisiae by microarray
  hybridization.
\newblock \emph{Molecular Biology of the Cell}, \textbf{9}, 3273--97.

\bibitem[{Spirtes \emph{et~al.}(2000)Spirtes, Glymour and Scheines}]{spirtes}
Spirtes, P., Glymour, C. and Scheines, R. (2000) \emph{Causation, {P}rediction,
  and {S}earch}.
\newblock The MIT Press, 2nd edition.

\bibitem[{St\"{a}dler and B\"{u}hlmann(2012)}]{missglasso2010}
St\"{a}dler, N. and B\"{u}hlmann, P. (2012) Missing values: sparse inverse
  covariance estimation and an extension to sparse regression.
\newblock \emph{Statistics and Computing}, \textbf{22}, 219--235.

\bibitem[{Thiesson \emph{et~al.}(2001)Thiesson, Meek and
  Heckerman}]{Thiesson2001}
Thiesson, B., Meek, C. and Heckerman, D. (2001) Accelerating {EM} for large
  databases.
\newblock \emph{Machine Learning}, \textbf{45}, 279--299.

\bibitem[{Tibshirani(1996)}]{tibshirani96regression}
Tibshirani, R. (1996) Regression shrinkage and selection via the {L}asso.
\newblock \emph{Journal of the Royal Statistical Society, Series B},
  \textbf{58}, 267--288.

\bibitem[{Troyanskaya \emph{et~al.}(2001)Troyanskaya, Cantor, Sherlock, Brown,
  Hastie, Tibshirani, Botstein and Altman}]{knnimpute2001}
Troyanskaya, O., Cantor, M., Sherlock, G., Brown, P., Hastie, T., Tibshirani,
  R., Botstein, D. and Altman, R. (2001) Missing value estimation methods for
  {DNA} microarrays.
\newblock \emph{Bioinformatics}, \textbf{17}, 520--525.

\bibitem[{Wille \emph{et~al.}(2004)Wille, Zimmermann, Vranova, F\"{u}rholz,
  Laule, Bleuler, Hennig, Prelic, von Rohr, Thiele, Zitzler, Gruissem and
  B\"{u}hlmann}]{wille}
Wille, A., Zimmermann, P., Vranova, E., F\"{u}rholz, A., Laule, O., Bleuler,
  S., Hennig, L., Prelic, A., von Rohr, P., Thiele, L., Zitzler, E., Gruissem,
  W. and B\"{u}hlmann, P. (2004) Sparse graphical {G}aussian modeling of the
  isoprenoid gene network in arabidopsis thaliana.
\newblock \emph{Genome Biology}, \textbf{5}.

\end{thebibliography}

\end{document}